\documentclass[fleqn,usenatbib]{mnras}

\usepackage{newtxtext,newtxmath}
\usepackage{soul}
\usepackage{float}
\usepackage[T1]{fontenc}
\usepackage[normalem]{ulem}
\DeclareRobustCommand{\VAN}[3]{#2}
\let\VANthebibliography\thebibliography
\def\thebibliography{\DeclareRobustCommand{\VAN}[3]{##3}\VANthebibliography}

\usepackage[dvipsnames]{xcolor}
\usepackage{multirow}
\usepackage{array}
\usepackage{booktabs}
\usepackage{graphicx}
\usepackage{subcaption}
\usepackage{amsmath}
\usepackage{amssymb}
\usepackage{xcolor}
\usepackage{changes}
\usepackage{adjustbox}
\usepackage{hyperref}
\usepackage{comment}
\usepackage{orcidlink}

\title[Hills EMRIs]{Extreme mass ratio inspirals from the Hills mechanism
\\
\Large The eccentricity distribution of Hills EMRIs in LISA}

\author[B. Sersante et al.]{
Biancamaria Sersante\orcidlink{0009-0008-2285-8370},$^{1}$\thanks{E-mail: sersante@strw.leidenuniv.nl}
Davide Mancieri\orcidlink{0009-0004-5106-9363},$^{2,3,4}$
Kevin Leopoldo Muñoz Zavala\orcidlink{0000-0001-7516-4665},$^{5}$
Luca Broggi\orcidlink{0000-0002-9076-1094},$^{3,4}$
\newauthor
Zephyr Penoyre\orcidlink{0000-0003-0244-4919},$^{1}$
Elena Maria Rossi\orcidlink{0000-0002-0587-6680},$^{1}$
Alberto Sesana\orcidlink{0000-0003-4961-1606}$^{3,4,6}$
\\
$^{1}$ Leiden Observatory, Leiden University, PO Box 9513, 2300 RA Leiden, The Netherlands\\
$^{2}$ Dipartimento di Fisica, Università degli Studi di Trento,
Via Sommarive 14, 38123 Povo, Italy\\
$^{3}$ Dipartimento di Fisica ``G. Occhialini'', Università degli Studi di Milano-Bicocca,
Piazza della Scienza 3, 20126 Milano, Italy\\
$^{4}$ INFN - Sezione di Milano-Bicocca,
Piazza della Scienza 3, 20126 Milano, Italy\\
$^{5}$ Institute of Science and Technology Austria (ISTA),
Am Campus 1, 3400 Klosterneuburg, Austria\\
$^{6}$ INAF - Osservatorio Astronomico di Brera,
Via Brera 20, 20121 Milano, Italy
}
\date{Received October 8, 2026. Accepted XX}
\pubyear{2026}

\begin{document}
\label{firstpage}
\pagerange{\pageref{firstpage}--\pageref{lastpage}}
\maketitle

%------------------------------------------------------------------

\begin{abstract}
Extreme mass ratio inspirals (EMRIs) are promising sources for future space-based gravitational-wave detectors. We investigate EMRIs produced through the Hills mechanism, in which a stellar-mass black-hole binary (BBH) is tidally separated by a massive black hole (MBH), leaving one component bound to it. 
We present the first self-consistent modelling of this channel, combining an orbit-averaged Fokker–Planck treatment of the nuclear star cluster (NSC), three-body modelling of the binary–MBH encounter, Monte Carlo evolution of captured black holes (BHs), and fully relativistic EMRI trajectories to follow the evolution from binary supply down to final plunge.
We find that Hills EMRIs are substantially more circular than classical EMRIs formed through two-body relaxation. For an MBH of $4 \times  10^{6}M_\odot$ and circular, equal-mass BBHs of $20 M_{\odot}$ with initial separation of 0.1 AU, EMRI mean eccentricities are $\approx 0.097$ at our adopted LISA-band entry threshold and $0.034$ at plunge, compared with $0.478$ and $0.228$, respectively, for classical EMRIs. 
Of the initial BBHs, $\approx 0.4$ per cent become EMRIs, whereas $\approx 60$ per cent of captured BHs do. For comparable NSC initial compact-object mass, we find  $\sim 1.6 \times 10^{-8} $ Hills EMRIs per year per galaxy, $\approx 12$ per cent of the classical EMRI yield. The Hills channel contribution to the overall eccentricity distribution depends strongly on the binary mass fraction in NSCs. We discuss implications for identifying EMRI formation channels with LISA and for quasi-periodic eruptions.
\end{abstract}


%%%%%%%%%%%%%%%%%%%%%%%%%%%%%%%%%%%%%%%%%%%%%%%%%%%%%%%%%%%%%%
\section{Introduction}
In recent decades, observations of the Galactic Centre have transformed our understanding of massive black holes (MBHs) and their environments. 
The first image of the event horizon of Sgr A* \citep{EHT2022}, together with 
orbital measurements \citep{Genzel2024} and multi-wavelength observations, directly confirmed 
general-relativistic predictions for the existence of MBHs and 
opened new avenues to explore accretion and jet physics. Yet many questions remain unanswered, in 
particular regarding the origin and assembly history of Sgr A* and the stellar and 
compact-object populations in its vicinity.

The complex dynamical environment of the Galactic Centre is central not only to test theories of gravity 
and accretion physics, but also to study hypervelocity stars \citep[HVS;][]{Hills1988}, stars ejected from the Galactic Centre at very high speed, and to improve our understanding of high-energy transients. 
These include tidal disruption events \citep[TDEs;][]{Hills1975, Luminet1986, 
Rees1988, StoneVasiliev2020, RossiStone2021} and extreme mass ratio inspirals \citep[EMRIs; see e.g.][]{AmaroSeoane2007, AmaroSeoane2022}. The latter are events in 
which a compact object orbiting an MBH spirals inward as it loses energy 
and angular momentum through gravitational-wave (GW) emission \citep{Peters1964, Misner1973, Maggiore2007}, ultimately merging with the MBH after tens of thousands of 
orbital cycles \citep{AmaroSeoane2018, RomSari2022}. Events in which the compact object sinks 
into the MBH on a nearly-radial trajectory or after only a few orbits are instead 
referred to as \emph{direct plunges} (DPs).

Two ingredients are required for EMRIs to exist: an MBH and compact objects in its vicinity. 
The first is well established: MBHs with masses $10^{6}$--$10^{10}\,M_\odot$ are 
commonly observed at galactic centres \citep{Magorrian1998, KormendyHo2013}, while 
lighter black hole (BHs) candidates ($10^4$--$10^5\,M_\odot$) have been identified in dwarf 
galaxies, although the occupation fraction is uncertain \citep{Greene2020}. 
MBHs typically reside within nuclear star clusters 
\citep[NSCs; see e.g.][]{Neumayer2020}: dense systems characterised by frequent dynamical 
interactions among stars, compact objects and the MBH itself \citep{BarOrKupi2013, Alexander2017}. This picture is supported by observations of HVSs \citep[e.g.][]{Koposov2020} and TDEs \citep[e.g.][]{Gezari2021}, which provide evidence that binary and single stars can reach distances small enough to strongly interact with an MBH; binaries get tidally separated to eject HVSs and populate stars in short-period bound orbits around MBHs \citep[][]{Verberne2025}, and single stars get tidally disrupted. 
 
EMRIs are key targets for future space-based GW detectors such as the Laser Interferometer Space Antenna (LISA)
\citep{Gair2017, AmaroSeoane2018, 2024arXiv240207571C}, Taiji \citep{Luo2021}, and 
TianQin \citep{Luo2025}, as their rates and orbital properties encode information 
about the dynamical environment of the MBH. For central MBHs with masses 
$10^5\,M_\odot \lesssim M \lesssim 10^7\,M_\odot$, GWs are emitted within the LISA band
%at frequencies  of $\sim 3\,\rm mHz$ during inspiral, well within the LISA band 
($10^{-4}$--$10^{-1}$\,Hz) \citep{Babak2007}. At plunge, most of the EMRI 
population has eccentricity $e_{\rm plunge}\in [0, 0.4]$, with a tail extending to 
$e_{\rm plunge} \approx 1$ and a peak near $e_{\rm plunge} \approx 0.2$ \citep{Mancieri2026}. 
The resulting long-lived ($\sim 10^4$ in-band cycles), harmonically rich GW signals 
make EMRIs highly distinctive sources.

The orbital properties and rates of EMRIs depend critically on their formation 
channel. 
The \emph{standard} dry mechanism is two-body 
relaxation \citep{SigurdssonRees1997, Hopman2005, Binney2009, AmaroSPreto2011, 
Merritt2013, Bar-OrAl2016A, Alexander2017, Broggi2022, Kaurperets2024, RomSari2025, Manciericliff2025}: 
repeated encounters with the surrounding stellar population diffuse the angular 
momentum of a compact object until it enters the GW-dominated regime and begins the slow inspiral 
characteristic of an EMRI.
The \emph{dry} scenario also includes models based on 
Kozai-Lidov oscillations in MBH binaries \citep[see e.g.][]{BodeWegg2014, 
NaozRose2022, Mazzolari2022, NaozHaiman2023} and supernova kicks 
\citep{BortolasMapelli2019}. 
Then, there is the so-called \emph{wet} scenario, which encompasses EMRIs formed through gas-assisted  capture and inspiral in AGN discs \citep[see e.g.][]{Levin2007, PanYang2021, 
PanLyu2021, DerdzinskiMayer2023}.

An alternative dry formation pathway involves the Hills mechanism \citep{Hills1988, Sari10, Kobayashi12, Rossi2014, Brown18}, with particular types of binaries:  a high-mass stellar or black-hole binary (BBH) arrives 
near the MBH on a highly eccentric, nearly parabolic orbit. The MBH's extreme tidal 
forces separate the binary, ejecting one component at high 
velocity and depositing the other on a bound orbit around the 
MBH. If the bound component is a BH (from a BBH), it can directly form an 
EMRI; if it is a high-mass star (from a stellar binary), it must first undergo stellar 
evolution.

The orbital properties predicted for the \emph{standard} two-body relaxation channel and the Hills one differ substantially. 
\citep{miller05} showed that Hills-induced EMRIs are expected to have pericentres 
of a few AU and apocentre distances of a few hundred AU, corresponding to moderate 
initial eccentricities ($e \sim 0.98$). Since Hills pericentres are much larger than 
the MBH gravitational radius, GW emission acts long enough to circularizes the orbit dramatically, and 
\citet{miller05} predicted eccentricities $e < 0.01$ upon entering the LISA band. 
\citet{raveh21} further showed that binary-capture products can produce 
low-eccentricity stellar BH EMRIs in mass-segregated clusters, though with somewhat 
higher eccentricities than the \citet{miller05} estimate, due to accounting for residual relaxation. In contrast, classical 
EMRIs formed through two-body relaxation typically begin with extreme eccentricities 
($e \gtrsim 0.9999$) and remain highly eccentric for most of their evolution. 
\citet{Hopman2005} found a characteristic eccentricity $e \sim 0.7$ at periods of 
$10^4\,\rm s$, corresponding to the LISA threshold frequency $f_\mathrm{LISA} = 
2\times10^{-4}\,\rm Hz$, while \citet{miller05} predicted $e \sim 0.5$--$0.9$ at LISA-band entry.

These eccentricity differences have direct observational consequences. Classical 
EMRIs, owing to their high typical eccentricity, produce complex GW waveforms with rich 
harmonic content that are challenging to model \citep{SperiGair2021, Speri2023, 
Speri2026}. Hills EMRIs, being nearly circular upon entering the LISA band, should 
produce more monochromatic and stable signals, potentially easier to detect and 
interpret.

An additional observational motivation to investigate different formation channels for EMRIs arises from quasi-periodic eruptions 
\citep[QPEs;][]{Miniutti2019, Giustini2020, Arcodia2021, Chakraborty2021, 
Quintin2023, Nicholl2024, Guolo2024, ArcodiaLiu2024, Arcodia2025, Chakraborty2025, 
HernGarcia2025, Bykov2025, Baldini2026}: recurring soft X-ray transients from MBHs 
in nearby galactic nuclei, with recurrence times of $\sim 2$--$10^2\,\rm hr$. 
Several physical mechanisms have been proposed to explain QPEs, including partial 
TDEs \citep{Coughlin2016, CoughlinNixon2020, NixonCoughlin2022} from highly 
eccentric white dwarfs \citep[see e.g.][]{Zalamea2010, King2020, King2022, 
Wang2022, Chen2023} or mildly eccentric stellar orbits \citep[see e.g.][]{Metzger2022, 
KrolikLinial2022, ZhaoZY2022, LinialSari2023QPE, LuQuataert2023}, and thermo-viscous 
instabilities \citep{Pan2022, Pan2023, KaurStoneQPE2023, Sniegowska2023}. The 
most-explored class of models instead invokes repeated crossings of the MBH accretion disc by 
a stellar-mass orbiter \citep[see e.g.][]{Xian2021, Linial2023, Franchini2023, 
TagawaHaiman2023, LinialMetzger2024b, LinialMetzger2024, VurmLinial2025, 
HuangLinial2025, Lam2025, Zhou2025, LinialMetzger2025, Yao2025, Pelle2026, 
Jankovic2026, Guo2026}. In this framework, the QPE light-curve properties are 
governed by the orbital parameters of the orbiting EMRI — including its inclination 
and apsidal/nodal precession \citep{Franchini2023, Chakraborty2025EMRIQPE, 
ZhouHuang2024, ZhouZhong2024, Allievi2026}. If this model applies, the distinct 
orbital distributions of Hills and classical EMRIs could, in principle, imprint 
different observational signatures on QPE light curves. QPEs would also act as electromagnetic precursors to EMRIs later detectable as LISA gravitational-wave sources \citep{2026arXiv260715360W}.

These motivations underlie the first self-consistent analysis of EMRI formation 
through the Hills channel, presented in this work. Different from \citet{miller05,raveh21}, our work provides a self-consistent treatment of the entire Hills-EMRI formation pathway, from the supply and tidal separation of binaries through the subsequent dynamical and GW-driven evolution to plunge. This allows us not only to predict the orbital properties of Hills EMRIs, but also to quantify their expected rates per galaxy and directly compare those with what expected from EMRIs formed through standard two-body relaxation. In summary, our goals are to provide:
\begin{enumerate}[I]
    \item a description of the orbital properties of Hills-mechanism EMRIs;
    \item a comparison of the eccentricity distributions between the Hills and 
    classical channels, at plunge and at the doorway of the LISA band;
    \item a comparison of the expected rate of EMRIs per galaxy from the two 
    channels.
\end{enumerate}

This paper is organised as follows. In Section \ref{steps} we describe our test case 
and the multi-step approach used to analyse Hills EMRI formation, introducing the 
initial conditions, assumptions, and numerical methods, and presenting intermediate 
results (result I). In Section \ref{results} we compare the orbital eccentricities 
of Hills to \emph{classical} EMRIs \citep[from][]{Mancieri2026}\footnote{Part of 
their data used in this work is publicly available at\\ 
\url{https://zenodo.org/records/17882705}.} at plunge and in the LISA band 
(result II), and provide estimates of the respective EMRI rates per galaxy (result III). 
In Section \ref{discussion} we discuss our results and their implications, including those for QPE observations.

%%%%%%%%%%%%%%%%%%%%%%%%%%%%%%%%%%%%%%%%%%%%%%%%%%%%%%%%%%%%%%
\section{EMRI formation in the Hills channel}\label{steps}

In this section, we describe our method for determining how many EMRIs can be produced through the Hills mechanism and their final eccentricity distribution, given an initial model for the NSC.

In this work, we consider an NSC similar to that of our own Galaxy, that harbors an MBH with mass  $M_{\bullet}$ of the order of  that of SgR A* i.e., $4\times10^6M_\odot$ 
\citep{Ghez2008, Gillessen2009, DoHees2019, Gravitycollab2024}. For simplicity we assume a non-spinning MBH. The central MBH, dominates the total potential inside its influence radius
\begin{equation}\label{r_inf}
    r_{\rm inf}=\dfrac{GM_{\bullet}}{\sigma^2}\sim2.22\rm \;pc\left(\frac{M_{\bullet}}{4\times10^6M_\odot}\right)\left(\frac{\sigma}{88 \; \rm km \;s^{-1}}\right)^{-2}.
\end{equation}
In Eq. (\ref{r_inf}) $G$ is the gravitational constant and $\sigma$ the velocity dispersion of stars in the bulge, related to $M_{\bullet}$ through the $M_{\bullet}-\sigma$ relation \citep[see][]{Gultekin2009}:
\begin{equation}
\sigma=88\; \mathrm{km}\;\mathrm{s}^{-1}\left(\frac{M_{\bullet}}{4\times10^6 M_{\odot}}\right)^{1 / 4.24} .
\end{equation}
The values of the other system's parameters are:
\begin{itemize}
    \item The total stellar mass in the NSC is $M_{\rm stars}=20 M_{\bullet}=8\times10^7 M_\odot$,
    \item The mass of each star is $m_s=1 M_{\odot}$,
    \item The total mass of BBHs is $M_{\mathrm{BBH}}=0.2 M_{\bullet}=8\times 10^5M_\odot$,
    \item The mass of each BBH is $m_{\mathrm{BBH}}=20 M_{\odot}$,
    \item The semi-major axis of each binary is $a_{\rm BBH}=0.1\;\rm AU$.
\end{itemize}
The NSC model includes $8\times 10^{7}$ solar-mass stars and $40,\!000$ BBHs, and we run 93 realizations\footnote{The number was chosen to roughly match the number of EMRIs produced by the standard channel in \citet{Mancieri2026}, the population we are comparing against.} of the model to generate more statistics.
We chose a semi-major axis of $0.1\;\rm AU$ since this is the minimum semi-major axis for the binary to avoid a GW-driven merger within a Hubble time. This estimate has been obtained by imposing the coalescence time \citep{Peters1964, Maggiore2007} for our circular {\it inner} BBH orbit,
\begin{equation}
    t_0=\frac{5}{256}\frac{c^5}{G^3}\frac{a_{\rm BBH}^4}{m_{\rm BBH}^2\mu_{\rm BBH}},
\end{equation}
with $\mu_{\rm BBH}$ the BBH reduced mass, to be equal to $10\;\rm Gyr$.  The tidal radius of the center of mass (CM) trajectory of the binary is
         \begin{equation}
        r_{\rm t}\simeq a_{\rm BBH}\left(\frac{M_{\bullet}}{m_{\rm BBH}}\right)^{1 / 3} \approx 5.85 \, {\rm AU}.
        \label{eq:rt}
    \end{equation} Both stars and BBHs distributions initially follow the same Dehnen profile with inner slope $\gamma=1.5$ \citep{Dehnen1993,Tremaine1994, Binney2009}
\begin{equation}\label{Denh}
n_i(r)=\frac{3}{8 \pi} N_i \frac{r_{\rm a}}{r^{3 / 2}\left(r+r_{\rm a}\right)^{5 / 2}} \theta\left(r-r_i\right) .
\end{equation}
Here $r_{\rm a}$ is the scaling radius at which the distribution changes slope from $r^{-3/2}$ to $r^{-4}$. To adequately scale $r_{\rm a}$ with the properties of the system, we set it to a multiple of the influence radius of the $\mathrm{MBH}$, $r_{\rm a}=4 r_{\rm inf}$ \citep{Merritt2013, Broggi2022}. This implies a density of $6\times 10^4M_\odot\rm{pc}^{-3}$ at the influence radius, consistent with observational estimates
for the Milky Way \citep{Schodel2007}.
The Heaviside function $\theta$ in Eq. (\ref{Denh}) is introduced to truncate the distribution at $r=r_i$. 
The quantity $r_i$ is a critical radius defining the size of the loss cone, which is the region of phase space corresponding to particles removed from the NSC (by tidal separation in case of a BBH, and by tidal disruption in case of a star). Specifically, an orbit is removed when it reaches a pericentre smaller than $r_i$, with $r_i$ the threshold radius that varies depending on the type of object under consideration.
In the case of stars $r_i=r_{\rm s}\left(\frac{M_\bullet}{m_{\rm s}}\right)^{1/3}$ is the tidal radius of the star\footnote{In this work we do not focus on the fate of the stellar population, considering it as a background scatterer; however, because it is important for the fate of BBHs, we do follow its phase space evolution self-consistently with the evolution of the BBH population.}, i.e., the radius at which the star is disrupted (and $r_{\rm s}$ the star radius), while for BBHs, $r_i=r_{\rm t}$. 
In terms of orbital elements, orbits penetrating the loss cone are those with a pericentre smaller than $r_i$, and particles on such orbits are removed from the NSC when they reach the pericentre.

The gravitational potential is assumed to be spherical and constant in time:
\begin{equation}
\Phi(r)=\frac{-GM_\bullet}{r}+2G(M_{\rm stars}+M_{\rm BBH})\left(1-\sqrt{\frac{r}{r+r_{\rm a}}}\right),
\end{equation}
where the second term, which we refer to as $\Phi_{\rm stars}$, is the potential generated by the extended star and BBHs Dehnen density distributions.
Indeed, although on time-scales longer than the relaxation time the potential might change, \cite{Broggi2024} argue that the variation in the MBH mass due to the accretion of stars and compact remnants can be neglected\footnote{We note that this is the case for our chosen MBH mass but not for smaller MBHs.}.

\begin{figure*}
    \centering
    \includegraphics[width=\textwidth]{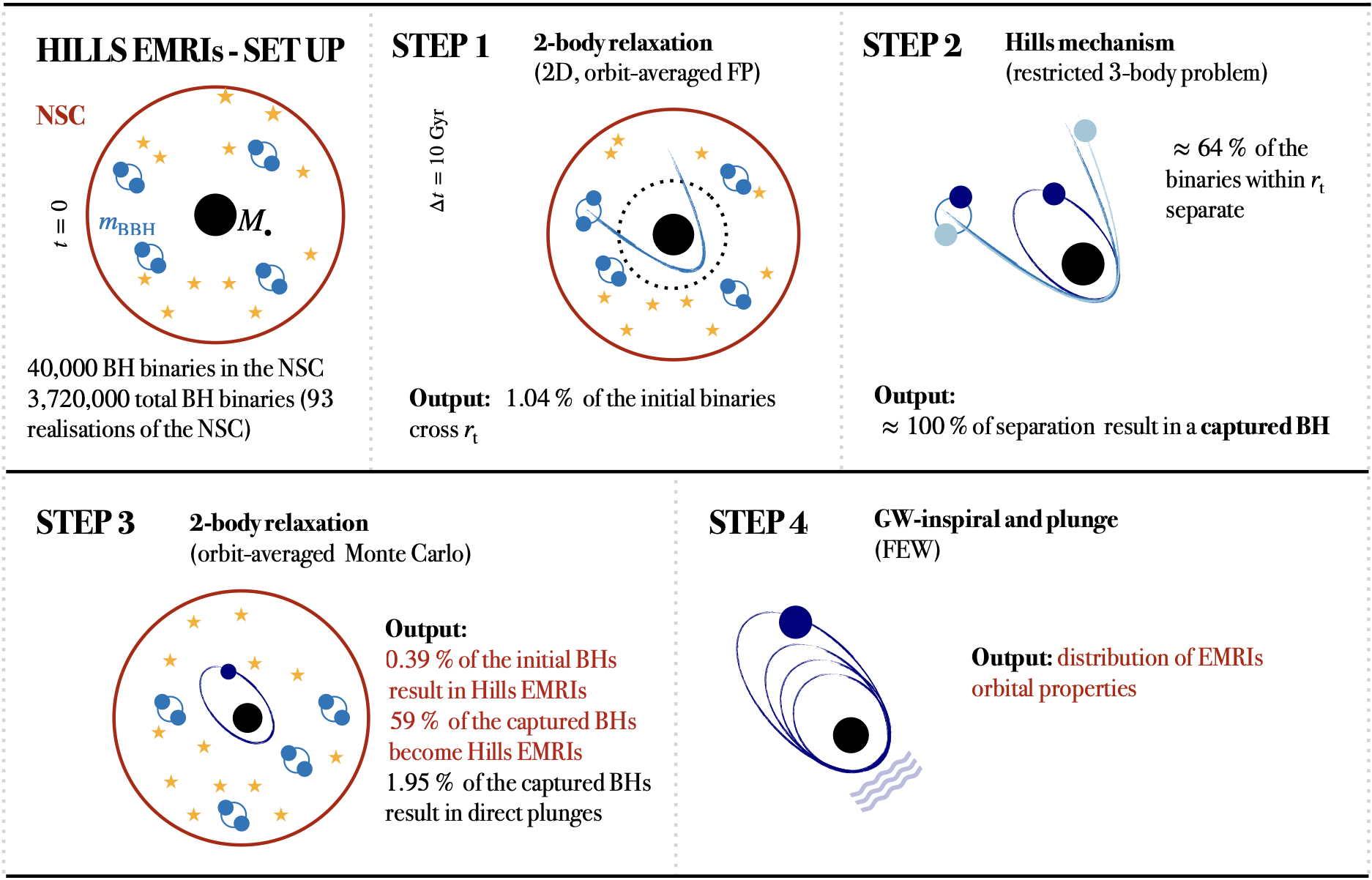}
    \caption{Cartoon to illustrate the various steps of our analysis of Hills EMRIs. At the end of step 1, binaries can either remain to populate the NSC or they can cross the tidal radius (the fraction of BBHs crossing $r_{\rm t}$ is $f^{r_{\rm t}}_{\rm BBH}=38,550/3,720,000=1.04$ per cent). Binaries that cross the tidal radius can either separate (in $64$ per cent of the cases) or survive ($\approx 36$ per cent of our systems). Binaries that separate result in a component being bound to the MBH and the other ejected. The bound component can be an EMRI, an SE or a DP. The total fraction of EMRIs produced by the initial population is $f^{\rm H}_{\rm EMRIs}=0.39$ per cent (see Section \ref{results}). The total fraction of captured BHs resulting in an EMRI is $59$ per cent. Systems that are EMRIs evolve, driven by GW-emission, until plunge. Their final properties are the result of Step 4.}
    \label{fig:sketch}
\end{figure*}

To analyse the formation of Hills EMRIs, we divide the calculation into four successive steps. The first three steps follow the formation and dynamical evolution of the captured compact objects, while the fourth follows the resulting EMRIs through inspiral to plunge. The first two steps -- concerning larger distances from the MBH -- are calculated in Newtonian dynamics, while the last two include modification to the dynamics due to gravitational wave emission. In the following, we provide a brief overview of the four steps of our approach, summarized visually in Figure \ref{fig:sketch}.

\begin{enumerate}
\item {\bf  Binary evolution through two-body relaxation.}

We first follow the evolution of the initial NSC under gravitational encounters between single-mass stars and BBHs. BBHs are treated as point-mass objects with twice the mass of a standard stellar-mass BH. Two-body relaxation gradually diffuses the orbital properties of the binaries, allowing a fraction of them to reach the vicinity of the MBH and eventually cross its tidal radius. The goal of this step is to characterize the orbital properties of the population of binaries that can enter the MBH loss cone. To do this, we solve the orbit-averaged Fokker-Planck (FP) equation in the two dimensions of energy and angular momentum. 
%This allows us to compute the time evolution of the distribution function of a spherical stellar system under two-body relaxation.
Our analysis is based on an improved version of the method developed by \cite{Broggi2022}.

\item{\bf Binary separation through the Hills mechanism.}

In the second step, we evolve the population of binaries entering the tidal radius through the Hills mechanism. We focus on binaries that are tidally separated by the MBH, leaving one component bound to it. The outcome of this step is a mapping between the orbital properties of the incoming binaries and those of the BHs captured following binary separation. We then determine whether each captured BH plunges directly into the MBH, which we define as a direct plunge (DP), or survives on an orbit bound to the MBH and can potentially evolve into an EMRI. We model this process using a semi-analytical treatment of the three-body problem \citep[following][]{Sari10,Kobayashi12,Brown18,SPR2025}.

\item{\bf Post-capture evolution.}

In the third step, we follow the evolution of the captured BHs under the combined effects of two-body relaxation and GW emission. The aim is to account for the residual effects of relaxation before the systems enter the regime in which their evolution is determined entirely by GW emission. Specifically, we determine how many captured BHs become EMRIs and characterize their orbital properties at the transition to deterministic GW-driven evolution. We model this stage using a Newtonian orbit-averaged Monte Carlo scheme, computing the GW energy and angular-momentum dissipation with the ``semi-relativistic'' approach of \citet{Hopman2005} (see Section \ref{step3section} for the detailed explanation). 
\begin{comment}
In this approach, the pericentre and apocentre are determined as the turning points at which the GR energy at a given angular momentum equals the effective GR potential. These relativistic turning points are then used to define Newtonian orbital elements, which are subsequently inserted into the classical flux formulae of \citet{Peters1964}. \citet{2006ApJ...639..999G} showed that this approach reproduces fully relativistic flux calculations significantly better than using the Peters equations with purely Newtonian orbital elements.
\end{comment}

\item {\bf EMRI evolution to plunge.}

Finally, for systems that become EMRIs, we follow their evolution through GW-driven inspiral to plunge in order to predict their final eccentricity distributions. We use {\it FastEMRIWaveforms} (FEW), a modular package that efficiently integrates EMRI orbits while accurately capturing the strong-field dynamics, where relativistic effects are essential \citep{FewChua2021,FewKatz2021,Speri2023,Chapman-Bird2025,chapman_bird_2025_15630565}.

\end{enumerate}

 The notation used throughout the paper consists of a superscript indicating the channel ($\rm H$ for Hills and $\rm C$ for classical), a superscript for the Hills channel indicating the evolutionary step ($r_t$ for the moment when BBHs cross the tidal radius and $\rm TSP$ for the moment when they tidally separate), and a subscript indicating the type of system under consideration ($\rm BH$ for single BHs, $\rm cap $ for captured BHs, $\rm BBH$ for binaries, $\rm EMRIs $ for EMRIs, and $\rm DPs $ for direct plunges). In the following sections, we detail each step separately.

\subsection{\emph{Step 1. Deposition of BBHs within the loss cone}}\label{FPstep}
To study the gravitational encounters that can lead BBHs in the proximity of the MBH, 
we employ and improve upon the FP
approach developed by \cite{Broggi2022}, as described in the following\footnote{A more thorough description will be the subject of a future publication.}. We consider a population of stars and BBHs\footnote{We do not include single BHs in the background population of scatterers. Single BHs of mass $10M_\odot$ would not behave significantly different from our binaries. In fact, in the simulation BBHs are considered as single BHs of double the mass, but with a different loss cone boundary.
%This assumption relies on the expectation that stellar-mass BHs undergo efficient mass segregation toward the innermost regions of the NSC. As a result, at the radii relevant for the relaxation processes considered here, gravitational encounters are expected to be dominated by stars, while the contribution of SBHs remains subdominant.
} and follow their evolution in energy and angular momentum space to analyse the flow of BBHs that crosses $r_{\rm t}$.\\

As gravitational encounters can be considered stochastic processes, they can be described analysing the evolution of the distribution function (DF) $g^i(\vec{x}, \vec{v},t)$ for $i\in\{\mathrm{stars}, \mathrm{BBHs}\}$ of each component present in the model; $g^i(\vec{x}, \vec{v},t){\rm d}^3\vec{x}{\rm d}^3\vec{v}$ represents the number of objects  in the six-dimensional phase-space volume ${\rm d}^3\vec{x}{\rm d}^3\vec{v}$ around $(\vec{x},\vec{v})$, at a certain time $t$. Assuming the potential of the star cluster to be spherical (including the MBH and the stellar potential)
and to evolve on time-scales longer than the typical orbital period, then the $g^i(\vec{x}, \vec{v},t)$ at initial time $t_0$ only depends on the integrals of
motion as $g^i(\vec{x}, \vec{v},t){\rm d}^3\vec{x}{\rm d}^3\vec{v}=f^i (\mathcal{E}, \mathcal{J})d\mathcal{E} d\mathcal{J}$, with $\mathcal{E}$ the energy per unit mass and $\mathcal{J}$ the angular momentum per unit mass.

The FP equation is solved numerically on a grid that is linearly spaced in
$\mathcal{R}\equiv(\mathcal{J}/\mathcal{J}_{\rm c})^2$, the square of the angular momentum normalised to the angular momentum of the circular orbit at the same energy, and uniformly spaced in
\[
s\equiv\log\left(1+\frac{\mathcal{E}}{\mathcal{E}_0}\right),
\]
where $\mathcal{E}_0=\sigma^2/5$ \citep[see also][]{Broggi2022}.
We adopt a resolution of $200\times200$ grid points,
with $s\in [ 0,s(E=E_{\rm max})]$, where $E_{\rm max}$ is the energy corresponding to a circular radius of $8 R_{\rm g}$, and $\mathcal{R}\in[0,1]$.
This choices ensures numerical convergence while keeping the computational cost manageable.

We impose the following boundary conditions on the BBH population; analogous conditions are applied to the stellar population:
\begin{itemize}
    \item we impose a zero-flux boundary condition at $s=0$, corresponding to the transition to unbound orbits, to restrict the distribution to gravitationally bound orbits in the total potential of the stars and the MBH (i.e., we do not consider orbits that are unbound from the NSC).
    
    \item We impose zero flux in $\mathcal{R}$ at $\mathcal{R}=1$, corresponding to circular orbits. Since $\mathcal{R}=1$ is the maximum physically allowed value for a given energy, no flux can pass beyond this boundary.
    
    \item The loss of objects through the loss cone is incorporated directly into the FP equation as a local sink term. This gives the total flux entering the loss cone

\begin{equation}\label{dsrrate}
\mathcal{F}_{\mathcal{R}}^{\mathrm{LC}}=4 \pi^2 L_c^2 \mathcal{R}_{\mathrm{LC}} f_{\mathrm{LC}} \sqrt{q} I_1(2 / \sqrt{q}) / I_0(2 / \sqrt{q})
\end{equation}
\begin{comment}
\begin{equation}\label{dsrrate}
\frac{\mathrm{d} N}{\mathrm{d}\mathcal{E}\mathrm{d}t}
=
4\pi^2\mathcal{J}^2_{\rm c}(\mathcal{E})
q(q^2+q^4)^{-0.25}
f(\mathcal{E},\mathcal{R}_{\rm lc}(\mathcal{E}),t).
\end{equation}
\end{comment}
Here, $\mathcal{R}_{\rm lc}$ is the value of $\mathcal{R}$ at the loss-cone boundary, corresponding to the tidal radius $r_{\rm t}$ and 
$q$ is the loss-cone diffusion parameter, defined as
\begin{equation}\label{q_def}
q=\frac{\mathcal{D}P}{\mathcal{R}_{\rm lc}}
=\left.\frac{P}{t_{\rm rlx}}\right|_{\mathcal{R}_{\rm lc}},
\end{equation}
where $\mathcal{D}$ is the energy diffusion coefficient of nearly-radial orbits (i.e. for $\mathcal{R}\rightarrow 0$), and $\mathcal{R}\mathcal{D}$ is the diffusion coefficient in angular momentum.
The first expression shows that $q$ measures the importance of angular-momentum diffusion over one orbital period for objects at the loss-cone boundary. Equivalently, using
$t_{\rm rlx}=\mathcal{R}/\mathcal{D}$,
$q$ can be interpreted as the ratio of the orbital period to the relaxation time-scale evaluated at the loss-cone boundary.
\end{itemize}
We note that the last point improves on the boundaries imposed in \cite{Broggi2022}. They implemented the loss of objects through the loss cone as a differential boundary condition at $\mathcal{R}=\mathcal{R}_{\rm lc}$, that becomes inaccurate for the extremely small values of $q$ that are typical of binary separations.

We integrate the FP equation for $10\,\rm Gyr$ and use the resulting flux of objects across $r_{\rm t}$ to construct the population of BBHs entering the tidal sphere. To map the orbital properties of these systems, an additional step is required.

\subsubsection{Orbital properties of BBHs crossing the tidal radius }
To quantify the orbital parameters of the flux of objects within $r_{\rm t}$, we start by saving adaptive snapshots in time ($\mathrm{d}t$) ensuring a maximum $3$ per cent relative variation in the DF across the grid. At each snapshot we save the quantities on the grid for both the stellar and the BBH population. Here, we treat the arrival of BBHs at $r_{\rm t}$ as a Poisson process; in each $\mathrm{d}t$, we first integrate the separation rate in Eq. (\ref{dsrrate}) over $\mathcal{E}$ and then we multiply it by the corresponding $\mathrm{d}t$. Multiplying the rate by \(dt\) gives the Poisson expectation value of the number of events in that time interval. The actual number of events in each interval is then drawn from a Poisson distribution with that expectation value. 

For each event, we use rejection sampling to assign the energy $\mathcal{E}_{\rm cm}$ (the subscript \emph{cm} refers to the binary CM and not to its inner orbit) of the binary according to the differential rate in energy at the time of the snapshot. After subtracting the stellar potential, we compute the semi-major axis of the CM trajectory around the MBH: $a_{\rm cm} = \frac{GM_\bullet}{2\mathcal{E}_{\rm cm}}$, where we adopt the convention that positive $\mathcal{E}$ corresponds to orbits bound to the MBH. The corresponding value of the loss-cone parameter, $q(\mathcal{E}_{\rm cm}$  ), is then evaluated for each event. Then, we  compute the diving factor 
\begin{equation}
  \beta=\frac{r_{\rm t}}{r_{\rm p,cm}},
\end{equation}
which quantifies how deeply the binary penetrates the MBH tidal sphere. Thus, $\beta<1$ corresponds to encounters outside the tidal sphere, while  $\beta >1$ corresponds to encounters penetrating within it. For each snapshot $\beta$ is drawn from the from the predicted distribution \citep[see][]{Broggi2025}
\begin{equation}
\frac{\mathrm{d}N}{\mathrm{d}\beta}=\tilde{N}
I_0\left(\frac{2}{\sqrt{q\beta}}\right)\beta^{-2},
\end{equation}
with $\beta\in[1,\infty)$, and normalisation $\tilde{N}= \left[\sqrt{q}I_1(2/\sqrt{q})\right]^{-1}$, where $I_k$ is the modified Bessel function of the first kind of order $k$.
From $(\mathcal{E}_{\rm cm},\beta)$, we reconstruct the orbital elements by first obtaining $r_{\rm p,cm}=r_{t}/\beta$
and then computing $(e_{\rm cm} = 1-r_{\rm p,cm}/a_{\rm cm})$. These orbital elements (in the potential of the MBH only) define the initial conditions for the subsequent evolution of the binaries in Step 2. We repeat this process for 93 different realizations and a total of $N^{\rm r_t}_{\rm BBH}=38,550$ binaries crossing $r_{\rm t}$.

\begin{figure*}
    \centering
    \includegraphics[width=\textwidth]{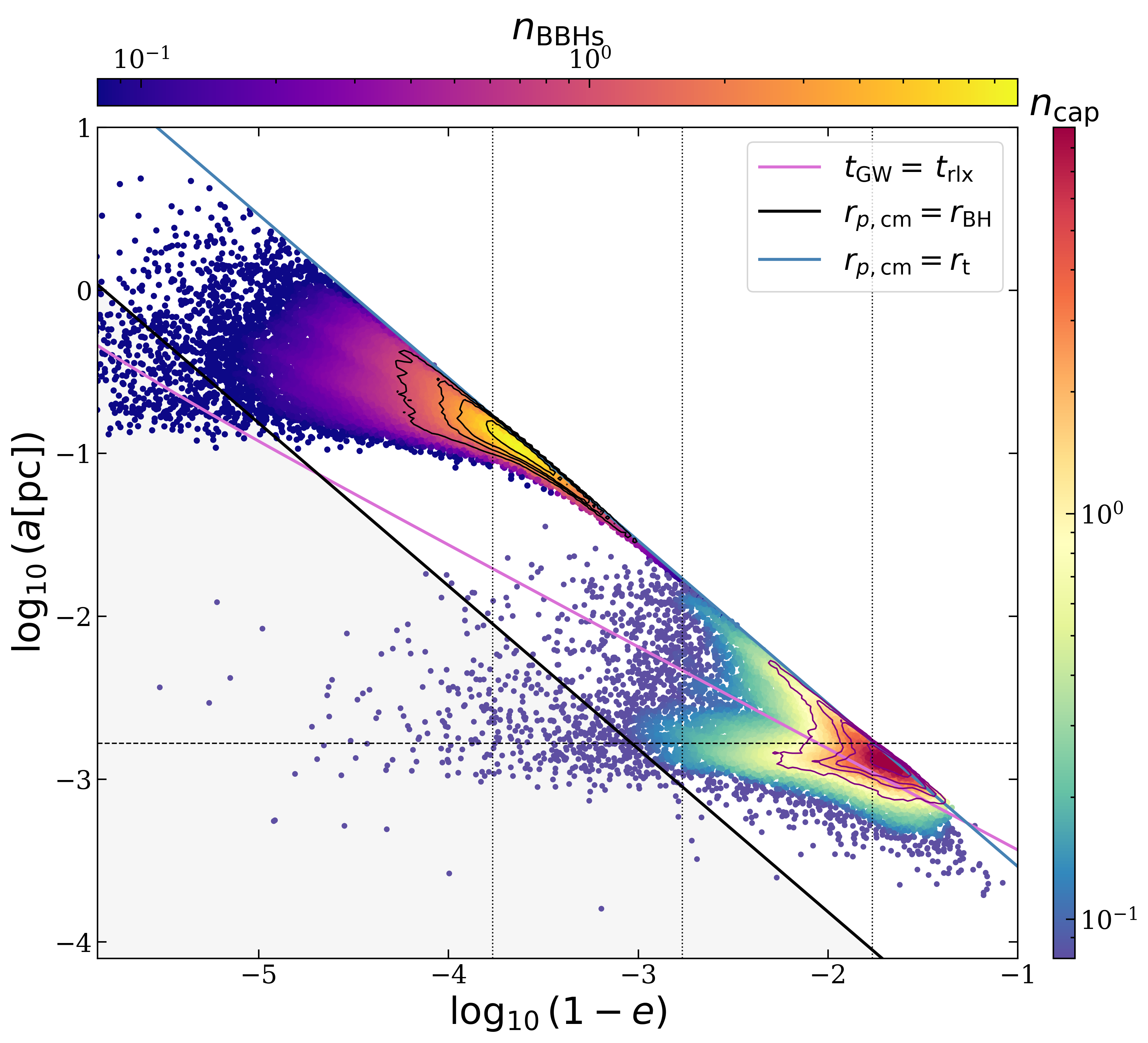}
    \caption{
    Scatter distributions of binaries that cross $r_{\rm t}$ (38550 systems) coloured by the number of BBH per logarithmic unit of semi-major axis and $(1-e)$, $n_{\rm BBH}$,
    %$n^{\rm H}_{\rm BBH}=\frac{dN}{d\log_{10}(a_{\rm cm})d\log_{10}(1-e_{\rm cm})}$ 
    (\emph{plasma} colormap), with $25,40,50,75-$ per cent contours (black solid lines).
    Scatter distribution of captured BHs (24592 systems), coloured by the number of BHs, per logarithmic unit of semi-major axis and $(1-e)$,
    %with number density $n_{\rm cap}=\frac{dN}{d\log_{10}(a_{\rm cap})d\log_{10}(1-e_{\rm cap})}$ 
    captured via Hills mechanism ($n_{\rm cap}$) (\emph{Spectral} colormap). The corresponding $25,40,50,75-$ per cent contours (purple) are lines of constant PDF. The black diagonal line represents the loss-cone boundary ($r_{\rm BH}=8R_g$) and orbital elements are here computed in the Newtonian framework. $r_{\rm BH}$ is the threshold in Newtonian phase space below which BHs plunge into the MBH. 
    The corresponding relativistic definition of the loss cone boundary can be found in Sec. \ref{step3section}
    The pink diagonal line marks $t_{\rm GW}=t_{\rm rlx}$ (at early times), computed as Eq. (\ref{tgw}) and Eq. (\ref{trlx}), respectively.
    The horizontal dashed line marks the characteristic semi-major axis for a separating binary $\alpha_D$, while the vertical dotted lines mark the characteristic change in eccentricities for separating binary, $\delta_D=1-e_{\rm D}$ for $\beta=1,10,100$ from right to left (as we are starting approximately on a parabolic trajectory, $\delta_D$ coincides with the final characteristic eccentricity value reached by the bound component upon binary separation).}
    \label{fig:total}
\end{figure*}

We show the binaries that cross the tidal separation radius in the semi-major-axis--eccentricity plane in Fig. (\ref{fig:total}), as the warm-coloured distribution.
In the plot, lines of constant pericentre satisfy $r_{\rm p,cm}=a_{\rm cm}\,(1-e_{\rm cm})=\rm const.$ and therefore appear as diagonal lines of slope $-1$ in this log plane. The sharp diagonal boundary of the distribution corresponds to
$r_{\rm p,cm}=r_{\rm t}$. 
We find that binaries that cross $r_{\rm t}$ occupy a confined region of parameter space, determined by the initial conditions. A negligible fraction of these is scattered directly within the BH loss cone; these systems are identified as DPs at the beginning of Step 3 of our analysis. The region enclosed by the $75$ per cent  contour contains the highest density of the  $(\log_{10}(a),\log_{10}(1-e))$ distribution, with semi-major axes between $\sim 0.1-0.2\;\rm pc$ and eccentricities between $\sim 0.9997- 0.9998$. 

The properties of binaries scattered within the tidal radius are presented  more extensively in Figure \ref{fig:betaecm}\footnote{The binaries are scattered in the close proximity of the MBH and, in the next step, are followed for only a single pericentre passage around the MBH. The stellar potential is therefore expected to be negligible, and the distributions shown in Fig. \ref{fig:betaecm} are computed assuming an MBH-dominated potential, ignoring the stellar contribution.}.

\begin{figure*}
        \centering
        \includegraphics[width=\textwidth]{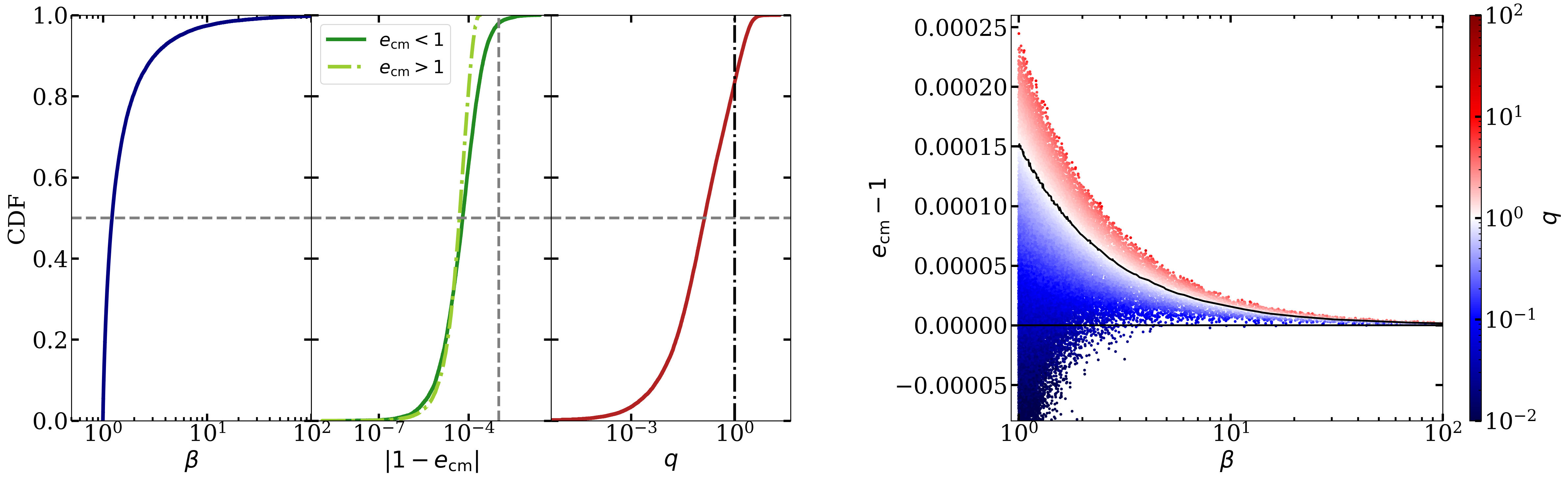}\caption{Orbital properties of the binaries reaching the tidal disruption radius, in the potential of the MBH only.
        \emph{Leftmost plot,first panel}: CDF of the diving factor $\beta$. \emph{Leftmost plot, second panel}: CDF of the eccentricity of the binary's CM, $e_{\rm cm}$, along its trajectory around the MBH. 
        The solid dark line (light dashed) is the CDF for binaries on orbits bound (unbound) to the MBH. The dashed vertical line marks $10^{-3}$ where the CDFs of bound and unbound orbits are, respectively, 0.98 and 1. 
        \emph{Leftmost plot, third panel}: CDF of the loss-cone diffusivity parameter $q$. The vertical dashed-dotted line corresponds to $q=1$, indicating the approximate boundary between the full and empty loss cone regime.
        In all panels, the gray horizontal dotted line marks the $50$ per cent probability. \emph{Rightmost plot}: Scatter plot of the CM-eccentricity vs $\beta$, coloured by $q$.
        The black solid line corresponds to $q=1$. We left out
        $0.27$ per cent of the systems, with $\beta>100$, for visualization purposes in the leftmost and rightmost panels.}
        \label{fig:betaecm}
        %\ContinuedFloat %      
    \end{figure*}

In the first panel of the leftmost plot, we show the cumulative distribution function (CDF) of the diving factor $\beta$ at the beginning of the three-body encounter. 
The majority of binaries enters the loss cone exactly when they reach the MBH's tidal radius ($\beta=1$) and shallow encounters are the most common: $80$ per cent of the systems have $\beta\leq2$, and $95$ per cent of the systems have $\beta\leq5$. The median value of $\beta$ is $1.22$. Only $2$ per cent of the systems have $\beta>10$, indicating that deeper encounters are comparatively rare.
In the second panel of the same plot, we show that the binaries approach the MBH on trajectories that are either slightly more or slightly less eccentric than a parabolic orbit. In particular, the CDF spans $|1-e_{\rm cm,0}|\in[10^{-7}-10^{-3}]$ corresponding to nearly parabolic orbits.
In the third, we show the CDF of the BBH loss-cone diffusion parameter $q$. We recall that the loss cone for BBHs is defined by their tidal radius $r_{\rm t}$. Equation  (\ref{q_def}) shows that  $q\gg 1$ corresponds to efficient angular-momentum relaxation: the angular momentum can change substantially over one orbital period, allowing systems to enter and leave the loss cone within a single orbit. In this regime, orbits penetrating the loss cone radius are continuously repopulated and the loss cone remains nearly full, with systems spending a fraction of their orbital periods  (i.e., $t_{\rm rlx}=P/q$) on such orbits. This is the full or pinhole loss-cone regime. Conversely, when $q \ll 1$, angular-momentum changes over one orbital period are small and relaxation is slow. Disruptive orbits are inefficiently repopulated, and the loss cone is effectively empty, corresponding to the empty or diffusive loss-cone regime. Here we refer specifically to the loss cone defined by the BBH tidal radius. 
%We find that $83.3$ per cent  of the systems have $q<1$ and are therefore expected to evolve predominantly in the empty-loss-cone regime, with a median value of $q=0.13$.
In the rightmost plot, we colour the  $(e_{\rm cm}-1) -\beta$ parameter space by $q$. We find that  $83.3$ per cent of the binaries, including both bound and unbound trajectories (with respect to the MBH), are produced in the empty-loss-cone regime, with a median value of $q=0.13$, while only a small fraction lies in the full-loss-cone regime, above the black line marking $q=1$. In the empty-loss-cone regime, binaries enter the loss cone slowly through two-body relaxation and therefore typically reach only the tidal radius, corresponding to $\beta\sim 1$. In contrast, the pericentre distribution is approximately flat from $0$ to $r_{\rm t}$, allowing some binaries to penetrate much deeper. This produces the high-$\beta$ tail. Only a very small fraction of binaries penetrate deeply into the tidal sphere, reaching pericentres of  $\approx 10^{-2} r_{\rm t}$, corresponding to $\beta > 10^2$; these systems are omitted from the figure for visual clarity.

The next step consists of evolving these BBHs through the Hills mechanism, starting from the distributions of $\beta$ and $e_{\rm cm}$ that define their CM trajectories.

\subsection{Step 2. Binary separation via Hills mechanism}

\begin{figure*}
    \centering
    \includegraphics[width=\textwidth]{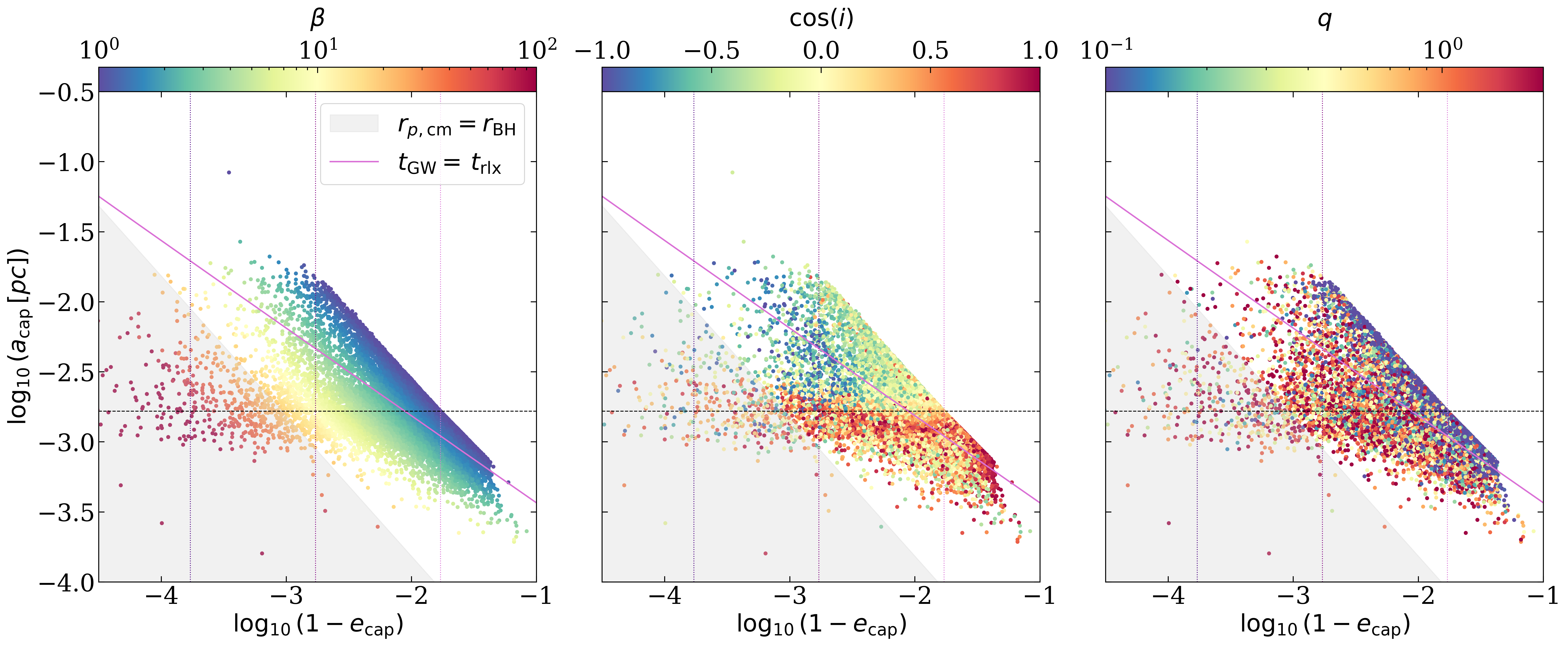}
    \caption{Scatter plot of the orbital properties (Newtonian semi-major axis and eccentricity) of the orbits of the Hills-captured BHs. 
    The points are coloured according to the corresponding initial $\beta$ (\emph{left panel}), to the cosine of initial inclination (\emph{central panel}) and by $q$ (\emph{right panel}).
    The pink solid line marks $t_{\rm GW}=t_{\rm rlx}$ (at early times) and the black diagonal line marks the boundary of the MBH loss cone, $r_{\rm BH}=8R_{\rm g}$. Horizontal and vertical lines are the same as in Fig. (\ref{fig:total})}.
     \ContinuedFloat %  
    \label{veryniceHills}
\end{figure*}

We now follow the BBHs that cross the tidal separation radius $r_{\rm t}$ and model their interaction with the central MBH. Crossing $r_{\rm t}$ does not necessarily result in binary separation: a binary may survive the encounter, merge, or be tidally separated \citep{SPR2025}. Here we focus on the latter outcome, which can leave one of the binary components bound to the MBH and the other ejected from the nuclear region. The \emph{captured BHs} constitute the progenitor population of the Hills EMRIs studied in the following sections.

We model the binary–MBH interaction following \citet{Sari10,Kobayashi12,SPR2025}. The calculation is based on a restricted three-body approximation, appropriate for the large mass ratio between the MBH and the binary, ($M_\bullet/m_{\rm BBH}\simeq 2\times10^{5}$), and on the hierarchy ($a_{\rm BBH}\ll r_{\rm t}$). In this regime, the CM of the binary follows, to leading order, an unperturbed Keplerian orbit around the MBH, while the relative motion of the two binary components is evolved in the time-dependent tidal field generated by the MBH. 

We therefore first determine the CM trajectory from the orbital parameters obtained in Step 1, namely the CM eccentricity, $e_{\rm cm}$ and the diving factor $\beta$ (shown in Fig. \ref{fig:betaecm}). For each binary, the CM orbit is then integrated through one pericentre passage. The calculation is Newtonian, given that the typical scale $r_{\rm t} \approx 5.85\;\rm AU$ is much larger than the MBH Schwarzschild  radius of $\approx 0.08$ AU\footnote{In fact, we verified that the Newtonian orbital elements are close to the relativistic ones for all $\beta$ values determined in Step 1.}.

The inner binary is initially assumed to be circular, with $a_{\rm BBH}=0.1$ AU. We randomly sample the remaining orbital degrees of freedom: the inclination ($i$) is drawn isotropically, i.e., uniformly in $\cos i$, while the inner-binary phase, argument of periapsis, and longitude of the ascending node are sampled uniformly in $[0,2 \pi]$. The binary is initialized sufficiently far from the MBH that the tidal perturbation is small with respect to the binary binding energy, while still dominating over perturbations from the surrounding stellar population, and a few inner orbital periods is enough for the tides to become dominant. We then follow the binary through a single pericentre passage and classify the outcome as survival or tidal separation. The latter constitutes $\approx 64$ per cent of the total BBHs crossing $r_t$ \footnote{Accounting for two more passages could result in a boost of around $20$ per cent to the fraction of separated systems for $\beta \gtrsim 1$ \citep{SPR2025}.}. 

For separated binaries, we determine the orbital energy and angular momentum of each component with respect to the MBH immediately after the encounter. A separated binary can produce two captured/ejected components -- in the case of an initially bound/unbound CM orbit -- or an ejected component together with a captured one. In our simulations, out of the initial 38550 binaries, the overwhelming majority of separated binaries produces one \emph{captured} BH; only four double-ejection and three double-capture events occur in the full sample. 

The resulting orbital properties of the captured BHs -- which we find independent of the time the parent BBH is injected into the loss cone -- are shown in Fig. \ref{fig:total}. The captured BHs occupy a substantially smaller region in semi-major axis and a more circular region of parameter space than their progenitor binaries. The highest-density part of the distribution (within $75$ per cent) lies around $a_{\rm cap}\sim1 0^{-3}\ {\rm pc},\, e_{\rm cap}\sim0.98,$
with the full population spanning approximately $e_{\rm cap}\simeq0.94-0.9999$ and $a_{\rm cap}\simeq3.2\times10^{-4}-3.2\times10^{-3}\,{\rm pc}$. Their pericentres remain approximately equal to those of the progenitor CM orbits, as expected \citep{Binney2009,Merritt2013}.

The characteristic semi-major axis of the captured component can be estimated analytically. For an equal-mass binary, \citet{SPR2025} obtain
\begin{equation}
\alpha_D \simeq Q^{2/3}a_{\rm BBH}
\simeq1.66\times10^{-3}\;{\rm pc}
\left(\frac{Q}{2\times10^5}\right)^{2/3}
\left(\frac{a_{\rm BBH}}{0.1\;{\rm AU}}\right),
\end{equation}
where $Q=M_\bullet/m_{\rm BBH}$. The characteristic scale of the simulated population agrees well with this prediction. Likewise, the characteristic change in eccentricity can similarly be estimated from
\begin{equation}
\delta_D\equiv1-e_D
\simeq Q^{-1/3}\beta^{-1}
\simeq1.71\times10^{-2}\;\beta^{-1}
\left(\frac{Q}{2\times10^5}\right)^{-1/3}.
\end{equation}
Because the incoming CM orbits are nearly parabolic, $(1-e_D)$ approximately corresponds to the eccentricity of the captured component after the binary separation. The simulated distribution follows this scaling closely: encounters with larger $\beta$ produce captured BHs with smaller $(1-e_{\rm cap})$, i.e., more eccentric post-separation orbits.

Figure \ref{veryniceHills} shows how the captured orbital properties depend on the parameters of the encounter. The strongest dependence is on the penetration factor and on the relative orientation of the binary orbit with respect to the CM orbit. Deeper encounters deposit the captured BH closer to the MBH (i.e. to the single BH loss cone). For the adopted parameters, binaries with $\beta\simeq 18.5$, corresponding to $r_{\rm p,cm}\simeq r_{\rm BH}=8 R_g$, can place a component directly inside the BH capture loss cone and therefore produce an immediate direct plunge.

The inclination also affects the post-separation orbital properties. At fixed $\beta$, initially prograde binaries tend to produce more tightly bound and more circular captured orbits, whereas retrograde configurations preferentially populate larger semi-major axes and higher eccentricities. This behaviour is consistent with the dependence of the separation probability on the binary orientation found by \citet{SPR2025}: prograde configurations are more readily separated, while strongly retrograde binaries are more resistant to tidal separation.

Finally, the loss-cone diffusivity parameter (q) inherited from Step 1 provides information about how deeply the incoming binaries can penetrate the tidal sphere. Most binaries reaching $r_{\rm t}$ do so in the empty-loss-cone regime and therefore have $\beta$ close to unity. The rare binaries originating in the full-loss-cone regime can undergo much deeper encounters and populate the high-$\beta$ tail. Consequently, both empty- and full-loss-cone binaries can produce captured BHs, but the most eccentric captured orbits are preferentially associated with the deeply penetrating, high-$\beta$ encounters.

An important consequence of the separation process is therefore that the Hills mechanism does not simply map a nearly parabolic incoming binary into an equally eccentric captured BH orbit. Instead, the tidal encounter deposits the captured component on a much more tightly bound orbit, with a characteristic semi-major axis set by the binary separation and MBH-to-binary mass ratio, while its pericentre remains approximately that of the incoming binary. These post-separation orbital properties provide the initial conditions for Step 3: the relaxation- and GW-driven evolution analysed in Section 2.3.

\subsection{Step 3. Relaxation-dominated evolution of Hills-captured BHs}\label{step3section}
To determine which of the BHs captured following binary separation become EMRIs, we follow their subsequent evolution under the combined effects of two-body relaxation and GW emission. Not all captured BHs form EMRIs: depending on their orbital properties, they can either plunge directly into the MBH, evolve into an EMRI, or remain on slowly-evolving orbits. We therefore introduce the following criteria to classify the different outcomes, following \citet[][]{Hopman2005, Bar-OrAl2016A,raveh21, Qunbar2024, Manciericliff2025}.

We define a captured BH as an EMRI once GW emission strongly dominates over relaxation, namely when
  \begin{equation}\label{Gwregion}
        t_{\rm GW} < 10^{-3} \;t_{\rm rlx},
    \end{equation}
where $t_{\rm GW}$ and $t_{\rm rlx}$ are the GW-emission and relaxation time-scales, respectively. The factor $10^{-3}$ ensures that the system is safely in the GW-dominated regime and is therefore very unlikely to be scattered back into the relaxation-dominated region. Conversely, when
    
\begin{equation}\label{rlxregion}
      t_{\rm GW} > t_{\rm rlx}  ,
    \end{equation}
the orbital evolution is primarily stochastic and driven by two-body encounters. We identify a system as a direct plunge (DP) if the particle is at pericentre and its angular momentum satisfies the relativistic condition,

\begin{equation}\label{DPs}
    J<4cR_{\rm g},
\end{equation} 
i.e., the condition for a particle arriving from infinity on a parabolic orbit to have low enough angular momentum to be directly captured by the MBH in a Schwarzschild spacetime. In the above equation, $R_{\rm g} = GM_\bullet /c^2$, where $c$ is the speed of light in vacuum.

We evolve the captured BHs until they either enter the strongly GW-dominated regime defined by Eq. \ref{Gwregion}, satisfy the DP criterion in Eq. (\ref{DPs}), or reach a slowly evolving region. The latter is defined by the combined time-scale
\begin{equation}\label{tcomb}
  t_{\rm comb}=\left(\frac{1}{t_{\rm GW}}+\frac{1}{t_{\rm rlx}}\right)^{-1} =100\rm Myr.
\end{equation}

\begin{figure}
    \centering
    \includegraphics[width=\columnwidth]{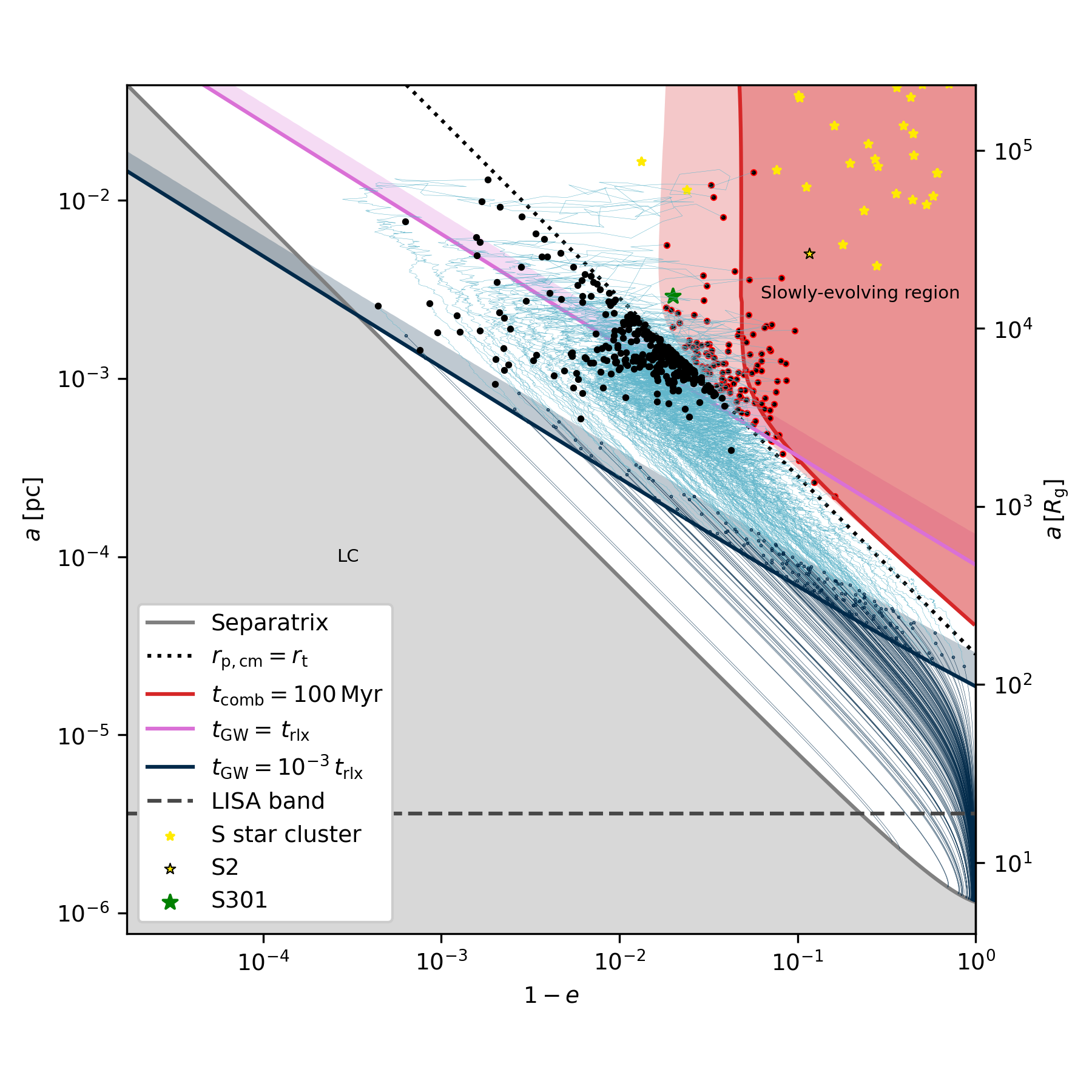}
    \caption{Examples of orbit evolution in the $(1-e,a)$ plane for a single realization (309 systems).
    DPs are not shown here. 
    The $t_{\rm GW}=t_{\rm rlx}$ pink line marks the boundary of the GW-dominated region\protect\footnotemark. 
    The tracks of captured systems before they become EMRIs (step 3) are marked in light blue. 
    Systems that cross the EMRI-detection boundary $t_{\rm GW}=10^{-3}t_{\rm rlx}$ (black line) are evolved in step 4 through inspiral until 
    plunge exclusively due to GW emission (dark blue, initially diagonal lines).
    The $t_{\rm comb} = 100\;\rm Myr$ red curve marks the boundary of the slowly-evolving region.
    The dashed horizontal line marks the semi-major axis corresponding to an EMRI orbital 
    period of $10^4\;\rm s$, an approximate indicator of entrance into the LISA band. 
    The dotted diagonal line marks pericentres corresponding to the tidal radius $r_{\rm t}$. 
    S-stars are plotted for reference 
    \citep[orbital solutions taken from][]{Gillessen2017, Verberne2025}, as well as the newly observed S301 \citep{S3012026}.}
    \label{fig:trajectories}

\end{figure}

Systems entering this region evolve only slowly on the time-scales relevant for our simulations, irrespective of whether relaxation or GW emission dominates. We therefore stop their evolution and classify them as slowly-evolving (SE) systems. These objects form a population of Hills-captured BHs on relatively tight, moderately eccentric orbits, analogous in this respect to the S-star population \citep{Verberne2025,Generozov2025}. The eccentricities shown in Fig. \ref{fig:trajectories} for these systems correspond to the stopping point and should not be interpreted as their present-day values: over longer time-scales, some may continue to evolve and circularize, while their semi-major axes are expected to change much more slowly until eccentricities reach $e\approx 0.5$, to then start changing on comparable time-scales.

\footnotetext{Since the diffusion coefficients used to compute the relaxation time-scale are those at the time of BBH separation, the stopping conditions for EMRIs and for SE systems vary from run to run (with a smaller spread for the former condition and a larger one for the latter). This is illustrated in Fig. \ref{fig:trajectories} through the shaded regions next to the $t_{\rm GW}=t_{\rm rlx}$, $t_{\rm GW}=10^{-3}t_{\rm rlx}$, and $t_{\rm comb}=100\;\rm Myr$ boundaries (at early times).}
    
We model the evolution of the captured BHs with an orbit-averaged\footnote{The Hills mechanism deposits BHs deep within the empty loss cone regime for single BHs (for which the LC edge is $r_{\rm BH}$), where the orbit-averaged approximation is accurate. We validated the orbit-averaged Monte Carlo approach here employed against the local approach from \citet{Manciericliff2025} for a subset of initial conditions, finding good agreement.}  Monte Carlo scheme that follows their energy and angular momentum under two-body relaxation and GW emission\footnote{A more thorough description of the code we used will be the subject of a future publication.}.  The initial conditions are the positions and velocities of the Hills-captured BHs obtained in Step 2. These are converted into energy and angular momentum using Newtonian prescriptions, with the energy subsequently corrected for the stellar potential, $\Phi_{\rm stars}$, which -- neglected in step 2 -- is reintroduced in this step. The evolution therefore starts from the time at which the corresponding BBH was extracted from the Fokker–Planck calculation in step 1, retaining the age of the NSC at the time of binary separation\footnote{Ignoring the time interval spent in step 2 is justified as separation happens on one pericentre passage, which is of the same order of one inner binary period.}.

At each Monte Carlo step, the diffusion coefficients are evaluated using the stellar distribution at the corresponding NSC age from the Fokker-Planck simulation of step 1. The relaxation time-scale is then computed directly from these diffusion coefficients as,
\begin{equation}
\begin{aligned}\label{trlx}
    t_{\mathrm{rlx}}&=\frac{\mathcal{J}^2}{\mathrm{D}^{2}_{\mathrm{o}}[\mathcal{J}]}\\
    &\approx 13.5\;\mathrm{Gyr}\;(1-e_{\rm cap}^2)\times\\
    &\hspace{4mm}\times \left(\frac{\sigma}{88 \mathrm{km\;s^{-1}}}\right)^{3}\left(\frac{\rho}{6\times 10^4 M_\odot \mathrm{pc}^{-3}}\right)^{-1}\left(\frac{m_{\rm s}}{M_\odot}\right)^{-1}\left(\frac{\ln\Lambda}{15}\right)^{-1},
\end{aligned}
\end{equation}
with $\mathrm{D}^{2}_{\mathrm{o}}[\mathcal{J}]$ the orbit-averaged, second-order diffusion coefficient in angular momentum, $\rho=6\times 10^4 M_\odot \mathrm{pc}^{-3}$ the density at $r_{\rm h}$ and $\ln\Lambda\approx\ln(M_\bullet/m_{\rm s})$ \citep{Binney2009, Merritt2013}.

The GW energy and angular-momentum fluxes ($\dot{J}_{\rm GW},\dot{E}_{\rm GW}$) are computed at each time step using the semi-relativistic prescription of \citet{Hopman2005}, in which the orbital turning points are determined in the relativistic potential and are then used to define the orbital elements entering the leading-order, orbit-averaged fluxes \citep[from][]{Peters1964}. This treatment provides a more accurate description of GW-driven evolution than applying the Peters equations directly to purely Newtonian orbital elements \citep{2006ApJ...639..999G}.
The time-scale of GW emission is computed as,
\begin{align}\label{tgw}
    t_{\rm GW}&=\min\left(\left|\frac{J}{\dot{J}_{\rm GW}}\right|,\left|\frac{E}{\dot{E}_{\rm GW}}\right|\right),\\ \nonumber
\end{align} i.e., as the shorter of the characteristic time-scales on which either energy or angular momentum changes by order itself due to GW-emission. Typical time-scales are of order of \citep{Manciericliff2025}
\begin{equation}\label{rlxtime_estimate}
\begin{aligned}
    t_{\rm GW} &\approx  9.18\times 10^3 \;\mathrm{Gyr}\;F(e_{\rm cap})\times\\
    &\hspace{3mm}\times\left(\frac{m_{\rm cap}}{10 M_\odot}\right)^{-1}\left(\frac{M_\bullet}{4\times 10^6M_\odot}\right)^{-2}\left(\frac{a_{\rm cap}}{10^{-3}\mathrm{pc}}\right)^{4},
\end{aligned}
\end{equation}
with 
\[F(e_{\rm cap})=\frac{(1-e_{\rm cap})(1-e_{\rm cap}^2)^{5/2}}{e_{\rm cap}(1+\frac{121}{304}e_{\rm cap}^2)},\]
such that, for the typical values of the various parameters in Eq. (\ref{rlxtime_estimate}), for example $ t_{\rm GW}(e_{\rm cap}=0.99) \approx 3.73 \mathrm{Myr}$ \citep[see e.g. ][]{ Maggiore2007, Merritt2013}.

The timestep $\Delta t$ is then selected adaptively such that the fractional changes in energy $E$ and $J$ due to both relaxation and GW-emission remain small during each step.
Stochastic kicks $\delta E$ and $\delta J$ are then drawn from Gaussian distributions and added (together with the systematic GW-driven drift) to the
current $(E,J)$. This procedure is iterated until the system becomes an EMRI, a plunge or hits the slowly evolving region. All time-scales ($t_\mathrm{rlx}$, $t_\mathrm{GW}$, $t_\mathrm{comb}$) are computed at the relevant NSC age using information from the Fokker-Planck simulation of step 1, meaning that the stopping conditions for identifying EMRIs and slowly-evolving systems change with the NSC age (see Fig. \ref{fig:trajectories}).

For systems satisfying the DP criterion, we account for the finite timestep by assigning a plunge probability equal to the fraction of the orbital period spent inside the loss cone, $\Delta t/P$ \citep{Broggi2025}. This accounts for the possibility that a system temporarily enters the loss cone without reaching pericentre during the timestep. In practice, this correction is negligible for our simulations because the systems predominantly evolve in the empty-loss-cone regime and the captured BHs have relatively small semi-major axes. Consequently, all DPs in our simulations are produced during Step 2, when the binary separation places the captured component directly inside the loss cone; subsequent relaxation does not produce additional DPs.

Figure \ref{fig:trajectories}, representative trajectories of Hills-captured BHs in the $(a,1-e)$ plane. Above the $t_{\rm GW}=t_{\rm rlx}$ boundary (pink line), the evolution is dominated by stochastic diffusion, which primarily changes the orbital angular momentum and hence the eccentricity. As GW emission becomes increasingly important below $t_{\rm GW}=t_{\rm rlx}$ , the systems begin to drift towards smaller semi-major axes. The trajectories remain at approximately constant pericentre while the orbits are highly eccentric. This behaviour can be understood from \citep{Peters1964},

\begin{equation}
    a(e)=c_0 \frac{e^{12 / 19}}{1-e^2}\left(1+\frac{121}{304} e^2\right)^{870 / 2299},
\end{equation}
which in the limit of $e \rightarrow 1$ becomes
\begin{equation}
        (1-e) a(e) \rightarrow \frac{c_0(425/304)^{870/2299}}{2} = {\rm const}.
\end{equation}
 Once the eccentricity becomes small, the pericentre starts to change and the trajectories do not conserve the pericentre anymore,
\begin{equation}
         a(e)\rightarrow c_0(e^{12/19}) \;\;\;\text{for} \;\;\;e\ll1 .
\end{equation}
 
When BHs cross the $t_{\rm GW}=10^{-3}t_{\rm rlx}$ boundary, we classify them as EMRIs and stop the Monte Carlo evolution. Their subsequent GW-dominated inspiral is followed separately in Step 4.

\begin{figure*} 
    \centering
    \includegraphics[width=\textwidth]{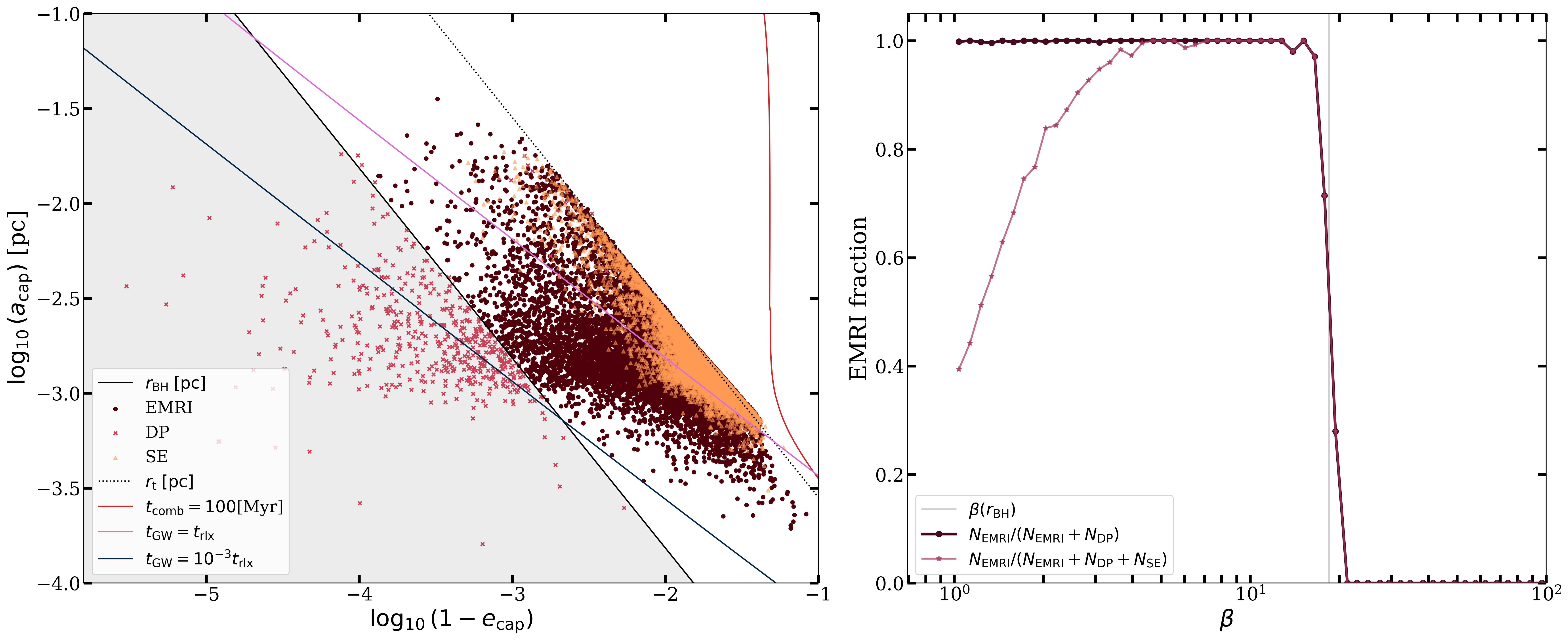}
    \caption{\emph{Left panel}: Scatter distribution of our captured BHs in the $(a_{\rm cap}-e_{\rm cap})$ parameter space coloured by their final fate (EMRIs: dark brown dots, DPs: pink crosses, SE systems: light orange triangles).
    Here $r_{\rm BH}=8R_g$ and orbital elements are computed from Newtonian prescriptions. \emph{Right panel}: Fraction of EMRIs as a function of the initial $\beta$. The darker line (circular markers) is the fraction of EMRIs relative to EMRIs and DPs, while the lighter one (star markers) is the fraction of EMRIs to EMRI, DPs, and SE systems. 
    The vertical black line marks $\beta=18.5$, corresponding to $r_{p,\rm{CM}}=r_{\rm BH}$.}
    \label{fig:p(beta)}
\end{figure*}

We identify $N_{\rm EMRI}^{\rm H}=14,517$ Hills EMRIs, corresponding to $\approx 59$ per cent of the captured BHs and $\approx 37.6$ per cent of the BBHs that cross the BBH tidal radius. Figure \ref{fig:p(beta)} summarizes the final outcomes in the $a_{\rm cap},e_{\rm cap}$ plane and shows that the initial encounter depth, quantified by the diving factor $\beta$, plays a key role in determining the fate of the captured BH. We discuss these dependencies and the resulting EMRI population in Section \ref{results}.

\subsection{Step 4. Gravitational-wave dominated evolution of Hills-captured BHs}\label{FEW}

We follow systems that cross the  $t_{\rm GW}<10^{-3}t_{\rm rlx}$ boundary, at which point GW emission dominates their evolution and we classify them as EMRIs. We evolve these systems through inspiral using the \texttt{FastEMRIWaveform} (FEW) package, which provides an efficient and accurate treatment of the orbital evolution in the Kerr spacetime.

We initialise the FEW integrations with the orbital parameters obtained at the end of Step 3, $(a^{\rm f}_{\rm MC},e^{\rm f}_{\rm MC})$. Specifically, we convert the semi-major axis to the semi-latus rectum according to $p^0_{\rm FEW} = a^{\rm f}_{\rm MC}[1-(e^{\rm f}_{\rm MC})^2]$ and set $e^0_{\rm FEW} = e^{\rm f}_{\rm MC}$. These parameters correspond to the dark-blue points in Fig. \ref{fig:trajectories}, close to the  boundary $t_{\rm GW}<10^{-3}t_{\rm rlx}$, and are directly suitable as input to FEW because they are computed within the same relativistic framework (as turning points of the relativistic potential).

During the inspiral, FEW evolves the orbital parameters according to the GW fluxes of energy and angular momentum. We use the \emph{KerrEccEqFlux}  module, which interpolates precomputed numerical fluxes obtained by solving the Teukolsky equation. These fluxes are available for eccentricities $e\lesssim 0.9$, although the precise upper limit depends on $p$ and can be as low as $e\simeq0.82$ \citep{Chapman-Bird2025}. For systems entering Step 4 with eccentricities larger than $0.9$, we therefore first use the PN5 module, based on analytical fluxes calculated to fifth post-Newtonian order \citep[see][]{Fujita2020}, and switch to \emph{KerrEccEqFlux} once the eccentricity falls within its range of validity. Further details of the method followed during this step are given in \cite{Mancieri2026}.

We terminate the evolution when the system crosses the Schwarzschild separatrix, $p = (6 + 2e) R_{\rm g}$, at which point the smaller BH plunges into the MBH. This relativistic criterion supersedes the approximate loss-cone boundary used in Step 3, valid only for parabolic orbits. The latter is appropriate for highly eccentric, nearly parabolic orbits, but becomes inaccurate as the orbit circularises during inspiral. The difference is small at large semi-major axes and high eccentricities, but becomes increasingly relevant for the tighter, more circular orbits reached near final plunge.

In Figure \ref{fig:trajectories}, we show that the orbital tracks, starting from $(a^{\rm f}_{\rm MC},e^{\rm f}_{\rm MC})$  (dark blue dots), transition to a coherent inspiral phase, as dark blue lines. Systems first reach semi-major axes of $3.6.\times10^{-6}\;\rm pc$ approximately marking the entrance of the LISA band (dashed horizontal line). This rough estimate relies on the assumption that the EMRI signal is approximately monochromatic (i.e., the orbit is circular) over a single orbital cycle, so that the LISA-band entry frequency, taken  indicatively to be $2 \times 10^{-4}$Hz, can be mapped to a characteristic period of
$10^{4}$s, which in turn corresponds to the above semi-major axis. Evolution then continues until the systems plunge. We will present results for the characteristic properties of Hills EMRIs in the next section.

\section{Results and comparison with classical EMRIs}\label{results}
We now present the main results of our simulations and compare Hills EMRIs with EMRIs formed through two-body relaxation. The results for the classical-EMRI population are taken from \cite{Mancieri2026}. We focus on two observables: the orbital eccentricity of EMRIs at plunge and at our adopted LISA-band entry threshold, and the relative rate of EMRIs produced by the two channels. Other formation channels, such as AGN-disc-assisted capture, are not considered here.

\subsection{EMRI orbital properties}
The eccentricity distribution is a key observable for distinguishing EMRI formation channels with future space-based gravitational-wave detectors. In particular, eccentricity affects both the harmonic structure and the duration of the signal in band: more eccentric EMRIs emit significant power in higher harmonics and generally require more cycles to circularize, whereas nearly circular EMRIs produce more monochromatic signals.

\begin{figure*}
    \centering
    \includegraphics[width=\textwidth]{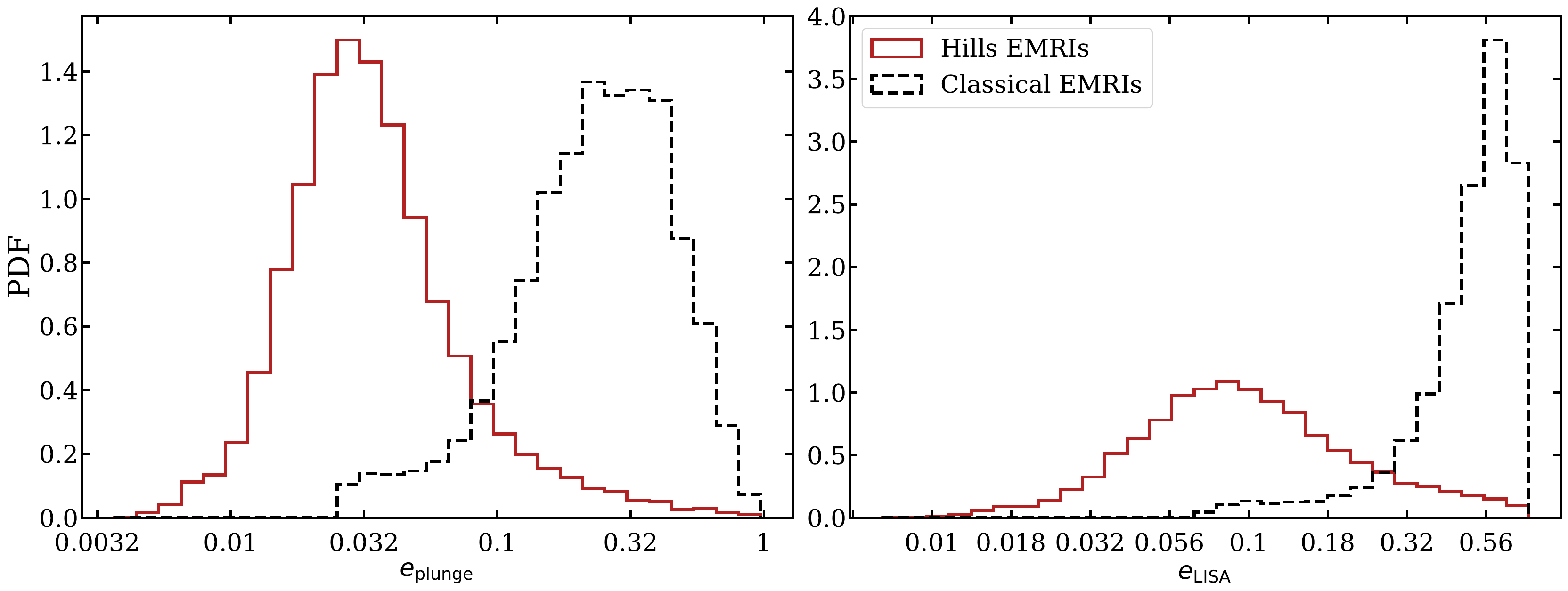}
     \caption{\emph{Left panel}:
    Probability density function (PDF) of the EMRI eccentricity distribution at plunge. 
     \emph{Right panel}: PDFs of EMRI eccentricity at the threshold of the LISA band.
     In both panels, the Hills distribution (solid red line) is compared with the \emph{classical}-EMRIs distribution from \citep{Mancieri2026} (black dashed line). The sharp cutoff of the classical distribution at $e^{\rm C}_{\rm plunge}=0.026$ likely results from having excluded the more circular systems that were diffused in the slowly-evolving region of the $(1-e,a)$ plane in \citet{Manciericliff2025}, since they employed a stricter condition to define that region than the one used in this work.}
     \label{fig:final ecc}
 \end{figure*}

Figure \ref{fig:final ecc}, shows the eccentricity distributions of Hills and classical EMRIs at plunge (left panel) and at our adopted LISA-band entry threshold (right panel). The two populations are clearly distinct. At plunge, the Hills distribution is strongly shifted towards lower eccentricities, with a mean eccentricity of $\bar{e}^{\rm H}_{\rm plunge} = 0.0346$, compared with $\bar{e}^{\rm C}_{\rm plunge} = 0.2282$ for classical EMRIs. The Hills distribution nevertheless extends over a broad range of eccentricities, including a tail towards high ones.

At the LISA-band entry threshold, the distinction remains pronounced. The Hills population has a mean eccentricity of $\bar{e}^{\rm H}_{\rm LISA} = 0.0975$, whereas the corresponding value for classical EMRIs is $\bar{e}^{\rm C}_{\rm LISA} = 0.478$. The Hills distribution is wider and spans approximately $0.0069 \lesssim e^{\rm H} _{\rm LISA}\lesssim 0.76$, while the classical distribution extends down to eccentricities only just below the peak of the Hills population. Thus, although both channels produce a range of eccentricities, Hills EMRIs are systematically more circular by the time they enter the LISA band. We note that the upper extent of both the Hills and classical eccentricity distributions $e\simeq 0.77$ in the right panel corresponds to the maximum eccentricity of a non-plunging orbit \citep[see][]{Hopman2005, Mancieri2026}, given a semi-major axis corresponding to the entrance into the LISA band at our adopted MBH mass ($3.6\times 10^{-6}$ pc).

Only $0.09$ per cent of Hills EMRIs have $e^{\rm H}_{\rm LISA} < 0.01$, while $2.83$ per cent reach $e^{\rm H}_{\rm plunge} < 0.01$. The fraction of events with eccentricities $< 0.01$ is therefore substantially smaller than predicted by the early analytical estimates of \cite{miller05}, which suggested that Hills EMRIs would enter the LISA band with almost-zero eccentricity. Our result is likely due both to the more sophisticated treatment of the whole evolution, which allows BHs to be deposited at smaller pericentres than assumed in the analytical estimates, and to residual dynamical relaxation, which prevents complete circularisation for most systems before they enter the LISA band \citep[as also seen by][]{raveh21}.

\begin{figure}
    \centering
    \includegraphics[width=\columnwidth]{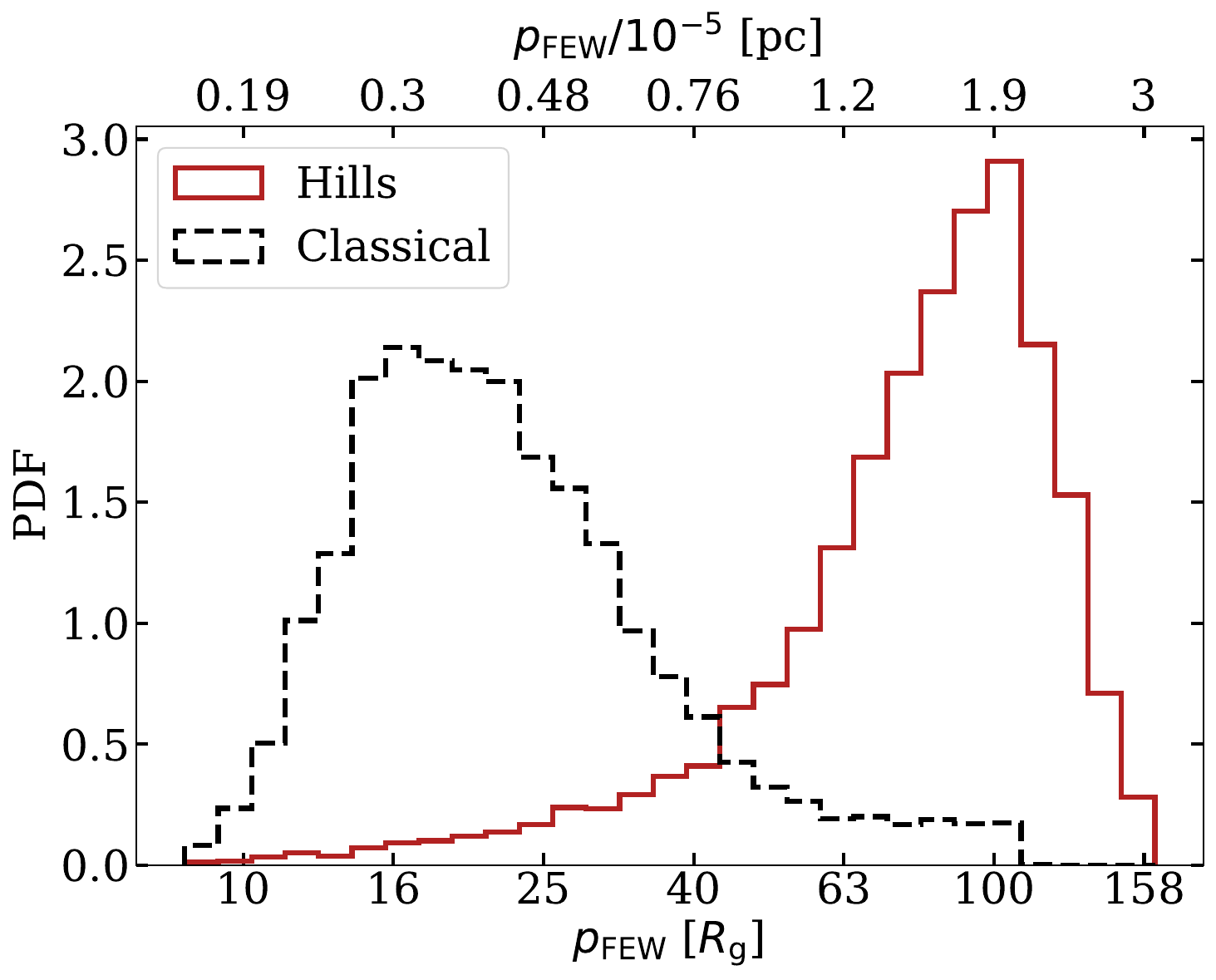}
    \caption{ Probability density function (PDF) of the semi-latus rectum $p$ of the EMRI orbits at the beginning of the evolution with FEW (Step 4). the Hills distribution (solid red line) is compared with the \emph{classical}-EMRIs distribution (data from \citet{Mancieri2026}) (black dashed line). The sharp cutoff in the classical distribution likely arises for the same reason as the cutoff in Fig.~\ref{fig:final ecc}.}
    \label{fig:pFEW}
\end{figure}

The origin of the difference between the two channels can be seen in Fig. \ref{fig:pFEW}, which shows the distribution of the semi-latus rectum p at the beginning of the GW-dominated evolution in FEW. Classical EMRIs enter this phase at smaller $p$, with a mean value of $\bar{p}^{\rm C }\simeq 4.89 \times 10^{-6} {\rm pc} =25.54\;R_{\rm g}$, whereas Hills 
EMRIs have a mean
$\bar{p}^{\rm H} \simeq 83.66 \; R_{\rm g}=1.60 \times10^{-5} {\rm pc}=$. Since $p$ is related to the specific angular momentum $J$, 
\begin{equation}
    p=\frac{J^2}{G M_{\bullet}},
\end{equation}
this means that Hills EMRIs enter the GW-dominated regime with larger angular momentum and farther from the loss cone than classical EMRIs.

This difference reflects the distinct formation histories of the two populations. Classical EMRIs are produced by relaxation-driven diffusion towards the loss cone and therefore decouple from the stellar environment relatively close to the separatrix. In contrast, Hills-captured BHs are deposited on tighter and more circular orbits than their progenitor binaries and can become GW dominated while still far away from the loss cone. They therefore have a longer GW-driven evolution over which gravitational radiation can circularize their orbits before the pericentre begins to shrink substantially.

The resulting orbital evolution naturally explains the eccentricity distributions in Fig. \ref{fig:final ecc}: Hills EMRIs enter the LISA band with substantially lower eccentricities than classical EMRIs and remain more circular throughout the subsequent inspiral. Their signals are consequently expected to be more nearly monochromatic, whereas classical EMRIs retain substantial eccentricity and therefore stronger harmonic structure. Hills EMRIs also spend more time in the LISA band, since they enter it, on average, farther from the separatrix.

The difference in the shapes of the two distributions alone, however, does not determine how informative a measured eccentricity is about the formation channel. The relative rate of EMRIs produced by the two mechanisms must also be taken into account.

\subsection{Implications for LISA observations}
We next compare the rate of EMRIs per galaxy produced by the Hills and classical channels for comparable initial populations of stellar-mass BHs. As a reminder, our reference Hills model contains $40,\!000$ BBHs of total mass $20 M_\odot$ per NSC, corresponding to the same total compact-object mass as $80,\!000$ single BHs of mass $10 M_\odot$ assumed in the classical model of \citet{Manciericliff2025}.

The 93 Hills realizations contain ${N}_{\rm BBH}=40,\!000\cdot 93=3,\!720,\!000$  initial binaries. Of these, $38,\!550$ cross the binary tidal radius over 10 Gyr, corresponding to $1.04$ per cent of the initial population. Approximately $64$ per cent of the binaries entering the tidal sphere are separated, producing $24,\!592$ captured BHs. Of these captured objects, $14,\!517$ become EMRIs, $480$ are direct plunges, and the remainder are classified as slowly evolving systems. The resulting fraction of the initial BBH population that becomes an EMRI is therefore
\begin{equation}
f^{\rm H}_{\rm EMRI}=\frac{N^{\rm H}_{\rm EMRI}}{N^{\rm TSP}_{\rm BBH}}\frac{N^{\rm TSP}_{\rm BBH}}{N^{r_{\rm t}}_{\rm BBH}}\frac{N^{r_{\rm t}}_{\rm BBH}}{N_{\rm BBH}}=\frac{14,\!517}{40,\!000 \cdot 93}=0.0039.
\end{equation}

Thus, approximately $0.39$ per cent of the initial BBHs in our reference model eventually become Hills EMRIs. In terms of the captured population, the corresponding fraction is much larger: $14,517/24,592 \simeq 59$ per cent. The bottleneck is therefore getting the BBH into the MBH tidal sphere, not the subsequent conversion of a captured BH into an EMRI.

\begin{figure}
    \centering
    \includegraphics[width=\columnwidth]{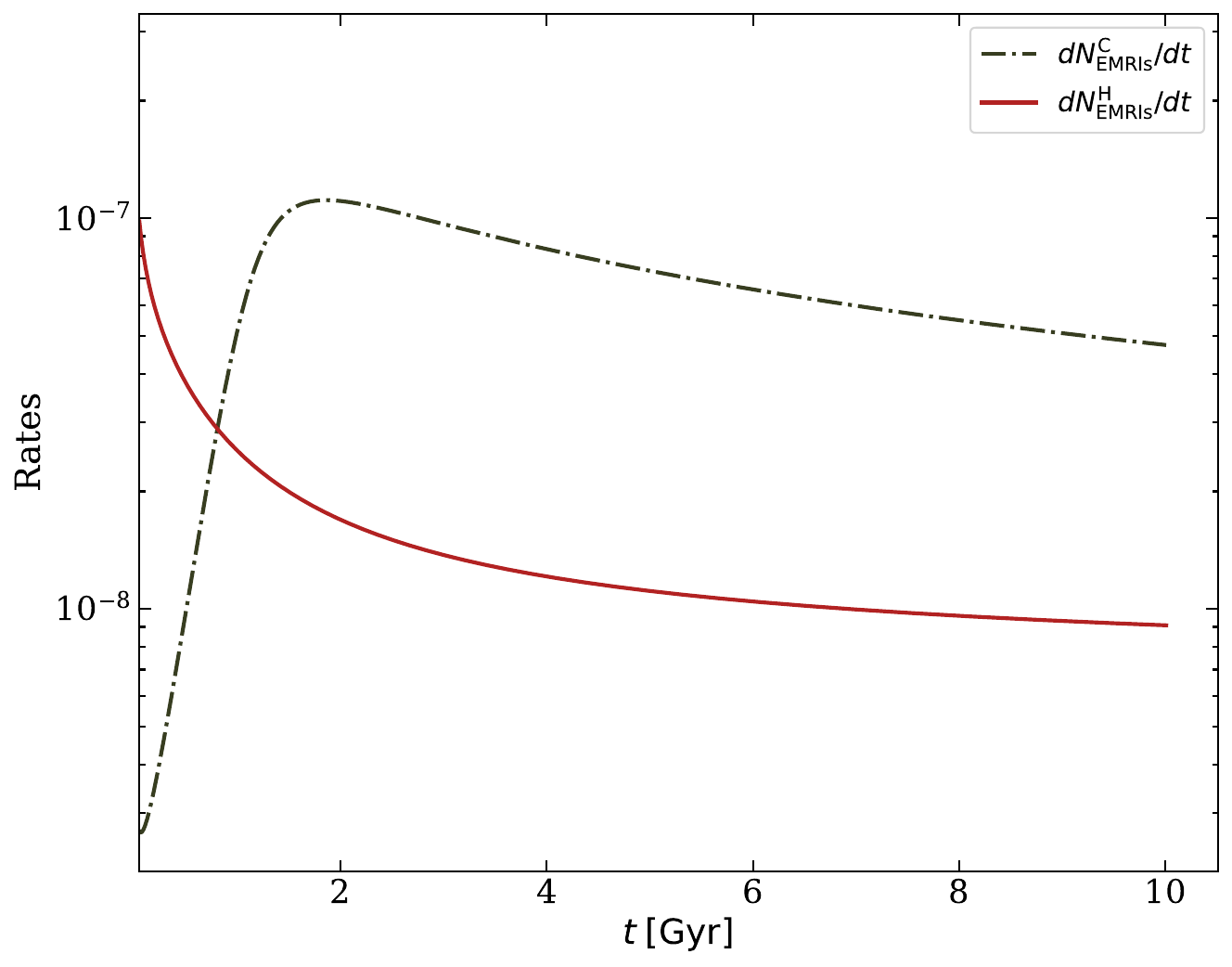}
    \caption{Rates for single BH EMRIs (black dashed line) and Hills EMRIs (red line) formation in
Step 1 of the analysis as a function of time.}\label{fig:rates}
\end{figure}

If the 93 realizations are interpreted as independent NSCs, each representing a different galaxy, the rate of Hills EMRIs produced per galaxy over 10 Gyr is
\begin{equation}
N^{\rm H}_{\rm EMRI, gal} =\frac{14,\!517}{93}=156;
\end{equation}
this number is equivalent to that obtained integrating the time-dependent EMRI production rate over the 10 Gyr evolution shown in Fig. \ref{fig:rates}.  

For comparison, the classical calculation starts with $80,\!000$ single $10 M_\odot$ BHs per galaxy and produces $N^{\rm C}_{\rm EMRI, gal}=1,\!344$ over the same period. This number is obtained by integrating in energy and time the differential rate used in \citet{Mancieri2026} to weight their $M_\bullet = 4 \times10^6 M_\odot$ runs. Thus,
\begin{equation}
{\frac{N^{\rm H}_{\rm EMRI, gal}}{N^{\rm C}_{\rm EMRI, gal}}=\frac{156}{1,\!344}\approx0.116}.
\end{equation}
For the  same total initial mass in stellar-mass BHs, the Hills channel therefore produces about $12$ per cent as many EMRIs as the classical channel, corresponding to approximately $88$ per cent fewer events. This comparison assumes the specific NSC model and BBH properties adopted here.

The relative contribution of the two channels in a realistic population additionally depends on the fraction $\mu$ of the total compact-object mass initially contained in binaries. We treat $\mu$ as a free parameter. For $\mu= 1/2$, $1/10$, and $1/100$, respectively, $50$ per cent, $10$ per cent, and $1$ per cent of the initial BH mass in the NSC is placed in binaries, with the remainder in single BHs.

\begin{figure*}
    \centering
    \includegraphics[width=\textwidth]{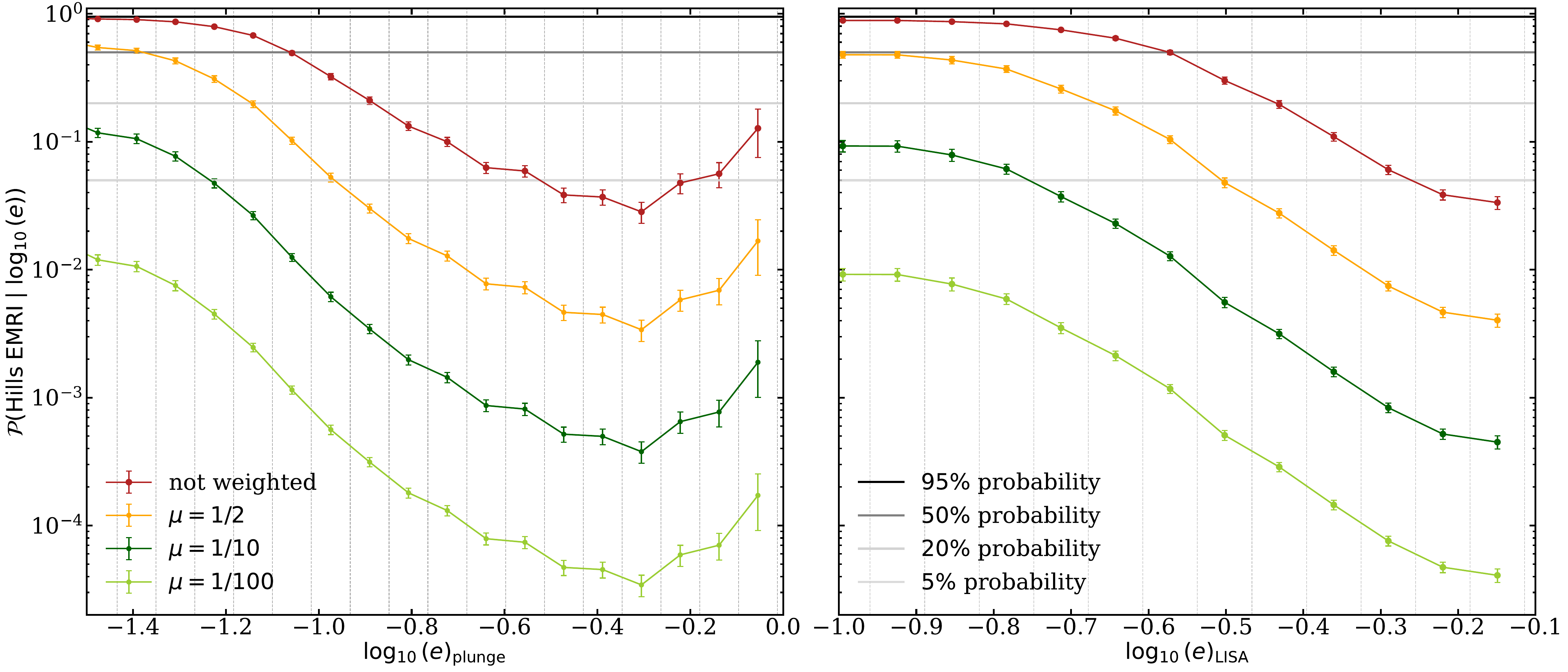}
\caption{Probability that an EMRI in a given eccentricity bin centred on value $e$, is an Hills EMRI as a function of eccentricity at plunge (\emph{left panel}) and at the entrance of the LISA band (\emph{right panel}).
The red upper line is obtained from our initial conditions: assuming the BBHs and single BHs have the same total mass. No further weighting is applied. 
The other curves are weighted by the product between the relative rate of Hills versus classical EMRIs per galaxy (i.e., $N^{\rm H}_{\rm EMRI}/N^{\rm C}_{\rm EMRI}=157/1344$) and by $\mu/(1-\mu)$.  The factor $\mu$ is the fraction of mass that is in binaries vs singles: $\mu=1/2, 1/10,1/100$ means that $50,10,1$ per cent, of the mass is contained in binaries and $50, 90,99$ per cent in singles, respectively.
The horizontal lines mark the $5,20,50$ per cent and $95$ per cent probabilities. Vertical dotted lines mark the edges of each bin.
The vertical error bars have been computed by propagation of uncertainty as explained in Appendix \ref{appendix}.}
\label{p(Hills|loge)}
\end{figure*}

For a given value of $\mu$, we weight the Hills and classical eccentricity distributions according to their relative rates per galaxy and compute the conditional probability that an EMRI with a measured eccentricity $e$ originated through the Hills channel,
\begin{equation}
  \mathcal{P}(\mathrm{Hills\; EMRI}|\log_{10}(e)\in \mathrm{bin})=\frac{H(\log_{10}(e))}{H(\log_{10}(e))+C(\log_{10}(e))},
\end{equation}
where $ H(\log_{10}(e))$ and $C(\log_{10}(e))$  are the corresponding weighted histogram densities. The weighting accounts for both the relative EMRI production efficiencies and the assumed binary mass fraction. The resulting probabilities are shown in Fig. \ref{p(Hills|loge)}, and the weighting procedure adopted is explained in its caption.

Without accounting for the relative production rates, the two eccentricity distributions imply that, at plunge, EMRIs with $e_{\rm plunge} \lesssim  0.087$ are more likely than not to belong to the Hills population under the reference model. Once the relative rates of the two channels and the binary fraction $\mu$ are included, this association becomes substantially weaker. For example, if $10$ per cent of the initial compact-object mass is in binaries (i.e., $\mu = 0.1$), the probability of a Hills origin remains below $20$ per cent even for the most circular systems at plunge. If $\mu = 0.01$, it remains below $5$ per cent across the eccentricity range.

At the entrance to the LISA band, the larger separation between the two eccentricity distributions makes low-eccentricity measurements more informative about the Hills channel. For $\mu = 1/2$, the conditional probability reaches $50$ per cent at approximately $e_{\rm LISA} \lesssim 0.12$ and stay above $20$ per cent for $e_{\rm LISA} \lesssim 0.25$. For $\mu = 1/10$, the corresponding low-eccentricity region has a conditional probability between $5$ per cent and $20$ per cent up to $e_{\rm LISA} \simeq 0.20$. These probabilities should be interpreted as population-level conditional probabilities under our assumed NSC model, binary fraction, and EMRI rates; they do not include LISA selection effects.

\begin{figure*}
    \centering
\includegraphics[width=\textwidth]{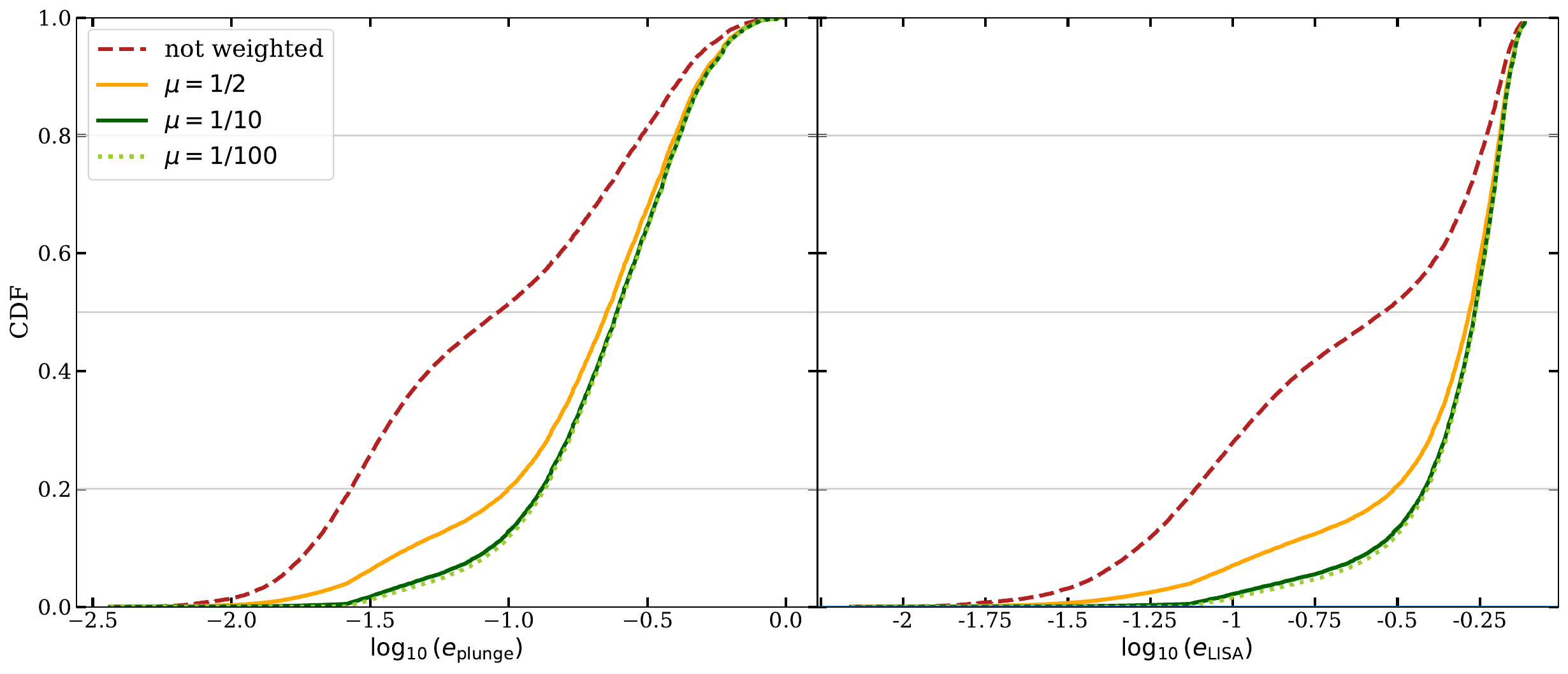}
\caption{CDF of detected eccentricity combining the two formation channels. \emph{Left panel}: eccentricity at plunge. \emph{Right panel}: eccentricity at the entrance of the LISA band.
The red dashed is obtained from our initial conditions: assuming the BBHs and single BHs have the same total mass and differ in number (with BBHs being 1/2 of the single BHs). No further weighting is applied. 
The other curves are weighted by the relative EMRI rate in the two channels and by the factor $\mu\in[1/2, 1/10,1/100]$, as explained in the caption of Fig. \ref{p(Hills|loge)}.
The horizontal lines mark the $20,50$ and $80$ per cent probabilities.
}
\label{fig:combinedCDF}
\end{figure*}

Finally, Fig. \ref{fig:combinedCDF} shows the combined cumulative eccentricity distribution of the two channels. This provides the complementary information needed to distinguish the probability of observing a given eccentricity from the probability that an EMRI with that eccentricity originated through the Hills mechanism. In other words, a low eccentricity can be informative about the formation channel only if such systems are sufficiently common in the underlying detectable population.

Together, Figs. \ref{fig:final ecc}–\ref{fig:combinedCDF} show that the Hills mechanism produces an EMRI population with systematically lower eccentricities than the classical channel, while the relative importance of the two populations depends strongly on the abundance of BBHs in galactic nuclei. The combination of an eccentricity measurement with an appropriate population model could therefore provide a statistical means of constraining the contribution of the Hills channel to the observed EMRI population. We notice that these considerations are based on a single MBH mass, which lies toward the high-mass end of the LISA detection range. Since EMRI detections are likely to be dominated by systems involving MBHs with $M \lesssim 10^{6}\,M_\odot$, a proper assessment of the statistical separability of the two formation channels will require extending our analysis to lower MBH masses.

\section{Discussion and conclusion}\label{discussion}

We have presented the first self-consistent modelling of EMRI formation through the Hills mechanism, following the evolution from the supply and tidal separation of BBHs to the subsequent dynamical and GW-driven evolution of the captured BH. We combine an orbit-averaged two-dimensional Fokker–Planck treatment of the NSC, restricted three-body calculations of the binary–MBH encounter, Monte Carlo modelling of two-body relaxation and GW emission, and fully relativistic evolution to final plunge with FEW. This approach allows us to connect the rate at which BBHs reach the MBH tidal radius to the orbital properties and final fate of the captured BHs.

For the specific NSC and BBH population considered here, we find that the outcome of the Hills encounter depends strongly on the initial pericentre of the binary. Binaries entering the tidal sphere close to the loss cone preferentially produce direct plunges, while systems with larger pericentres can produce slowly evolving objects or EMRIs. For our adopted population of equal-mass, circular BBHs with $m_{\rm BBH}=20\,M_\odot$ and $a_{\rm BBH}=0.1\mathrm{AU}$, EMRIs are produced over a broad range of encounter parameters, with the EMRI fraction increasing with decreasing $\beta$ up to the region where slowly evolving systems become important. These quantitative boundaries should not be regarded as universal, however, as they depend on the MBH mass and on the assumed BBH properties.

The main robust result of our calculation is the eccentricity distribution of Hills EMRIs. Hills EMRIs are substantially more circular than classical EMRIs both at plunge and when they enter the LISA band. In our reference model, the mean eccentricities are $\bar{e}^{\rm H}_{\rm plunge}=0.0346$ and $\bar{e}^{\rm H}_{\rm LISA}=0.0975$ for Hills EMRIs, compared with $0.2282$ and $0.477944$, respectively, for the classical population of \citet{Mancieri2026}. The difference arises from the orbital properties at the point where GW emission takes over from stellar dynamics. 
In fact, Hills-captured BHs are deposited on more circular orbits and with smaller semi-major axes. So, since the $t_{\rm relax}=t_{\rm GW}$ boundary is not parallel to the loss-cone boundary,
when they decouple from the nuclear cluster (i.e., when they enter the GW-dominated regime) they lie farther from the loss cone than classical EMRIs and occupy less eccentric orbits.
Since GW-driven evolution approximately preserves the pericentre while the orbit remains highly eccentric, these systems naturally enter the LISA band with lower eccentricities.
%Hills-captured BHs are deposited on more circular orbits, with smaller semi-major axes, and decouple from the nuclear cluster farther from the loss cone than classical EMRIs. They therefore start their GW-dominated evolution with lower eccentricities and have more time to circularize before the pericentre approaches the separatrix.

Our results refine the original prediction of \citet{miller05}, who argued that Hills EMRIs should enter the LISA band with $e \lesssim 0.01$. We find that only $0.09$ per cent of our Hills EMRIs have $e_{\rm LISA}<0.01$, although 2.83 per cent reach $e_{\rm plunge}<0.01$. Thus, while the Hills mechanism naturally produces a more circular population than the classical channel, most systems do not become perfectly circular before entering the LISA band due to a combination of the more sophisticated treatment of the whole evolution and of residual two-body relaxation preventing complete circularization. This result is broadly consistent with \citet{raveh21}, who followed the post-separation evolution of tidally captured compact objects and likewise found that relaxation can imprint a residual eccentricity on the EMRI population. We refrain from making a direct comparison with their results, as they report eccentricity distributions computed at orbital period of $10^3 \, \mathrm{s}$, which corresponds to a semi-latus rectum within the separatrix for an MBH of $4 \times 10^6 \, M_\odot$ \citep[see][]{Mancieri2026}.

The difference between the two channels has direct consequences for LISA observations. Hills EMRIs are expected to produce more nearly monochromatic signals, with less prominent harmonic structure than the typically highly eccentric classical population. They may therefore be easier to identify and model, although end-to-end EMRI search pipelines must be implemented and tested on realistic LISA data to verify this statement.
%a quantitative assessment of detectability requires folding the intrinsic population distributions through the LISA response and selection effects. 
%In particular, the eccentricity distributions found here should be combined with realistic LISA parameter-estimation studies to determine how well the two formation channels can be distinguished in practice.

We also find that the relative rate per galaxy of EMRIs produced by the two channels is important when interpreting an observed eccentricity. For the same total initial mass in stellar-mass BHs, our reference model produces approximately 156 Hills EMRIs per galaxy over 10 Gyr, i.e., an average rate of $1.6 \times 10^{-8}$ yr$^{-1}$ gal$^{-1}$, compared with $1344$ classical EMRIs, i.e., $1.3 \times 10^{-7}$ yr$^{-1} $ gal$^{-1}$. Thus, the Hills channel produces about $12$ per cent as many EMRIs, or equivalently about $88$ per cent fewer, under the assumptions adopted here. The relative contribution of the Hills channel depends strongly on the fraction $\mu$ of the compact-object mass initially contained in binaries. Consequently, an eccentricity measurement alone does not identify the formation channel: its interpretation depends on the relative abundance of the two populations. For our assumed distributions, low eccentricity can nevertheless provide statistical evidence for a Hills origin, particularly when the binary fraction is substantial ($\mu > 1/10$).

The Hills mechanism also produces ejected counterparts of the captured BHs. We find median ejection velocities of $v_{\rm D} \approx 3870$ km s$^{-1}$  for both the first and second components of the binary,  consistent with the analytical expected value \citep[see e.g. ,][]{SPR2025}. The ejected components therefore constitute a population of  $10 M_\odot$ BHs launched on unbound orbits from the Galactic Centre at exceptionally high velocities. 

All quantitative predictions in this work should be interpreted in the context of our deliberately simplified initial conditions. We consider a single MBH mass, a single stellar mass, and a BBH population with fixed mass, semi-major axis and initially circular orbits. A physically complete population model should instead explore the distributions of MBH mass, BBH masses, semi-major axes and eccentricities, as well as realistic binary fractions and their dependence on the properties of the NSC. Extending the calculation to lower MBH masses is particularly important for LISA, as the sensitivity to EMRIs is expected to peak around $M_\bullet =3\times 10^5\,M_\odot$ \citep{Babak2017}. These extensions will also be necessary to translate our per-galaxy rates into robust cosmological event-rate predictions.

Our treatment also excludes BBHs that survive the initial encounter and remain bound to the MBH as binaries. Such binary-EMRI systems (also known as b-EMRIs) require substantially harder binaries than those considered here \citep{Chen2018} and the probability for them to be in the EMRI region is very low (they are likely scattered awayif they are penetrating, or they simply remain part of the NSC binary population), and therefore represent a complementary channel rather than part of the population studied in this work. 
Similarly, we focus exclusively on BH EMRIs; stellar EMRIs produced through the Hills mechanism could have different orbital properties and may be relevant in other astrophysical contexts.

\subsection{Implications for Quasi-Periodic Eruptions}
 The possibility that QPE-producing stellar orbits could be generated through the Hills breakup of stellar binaries has been discussed by \citet{LuQuataert2023}, while \citet{LinialSari2023QPE} considered the subsequent GW-driven evolution and unstable mass transfer as a QPE mechanism. 
 Our results instead may provide a framework for assessing the Hills mechanism in models in which QPEs are produced by repeated crossings of an MBH accretion disc by an orbiting compact object; the observed recurrence time constrains the orbital period $P$, while the flare properties constrain the orbital eccentricity and inclination. The distinct ($P,e$) distributions of Hills and classical EMRIs therefore offer a way to test whether a Hills origin is compatible with the orbital parameters inferred for individual QPE sources.

\begin{figure}
    \centering
    \includegraphics[width=\columnwidth]{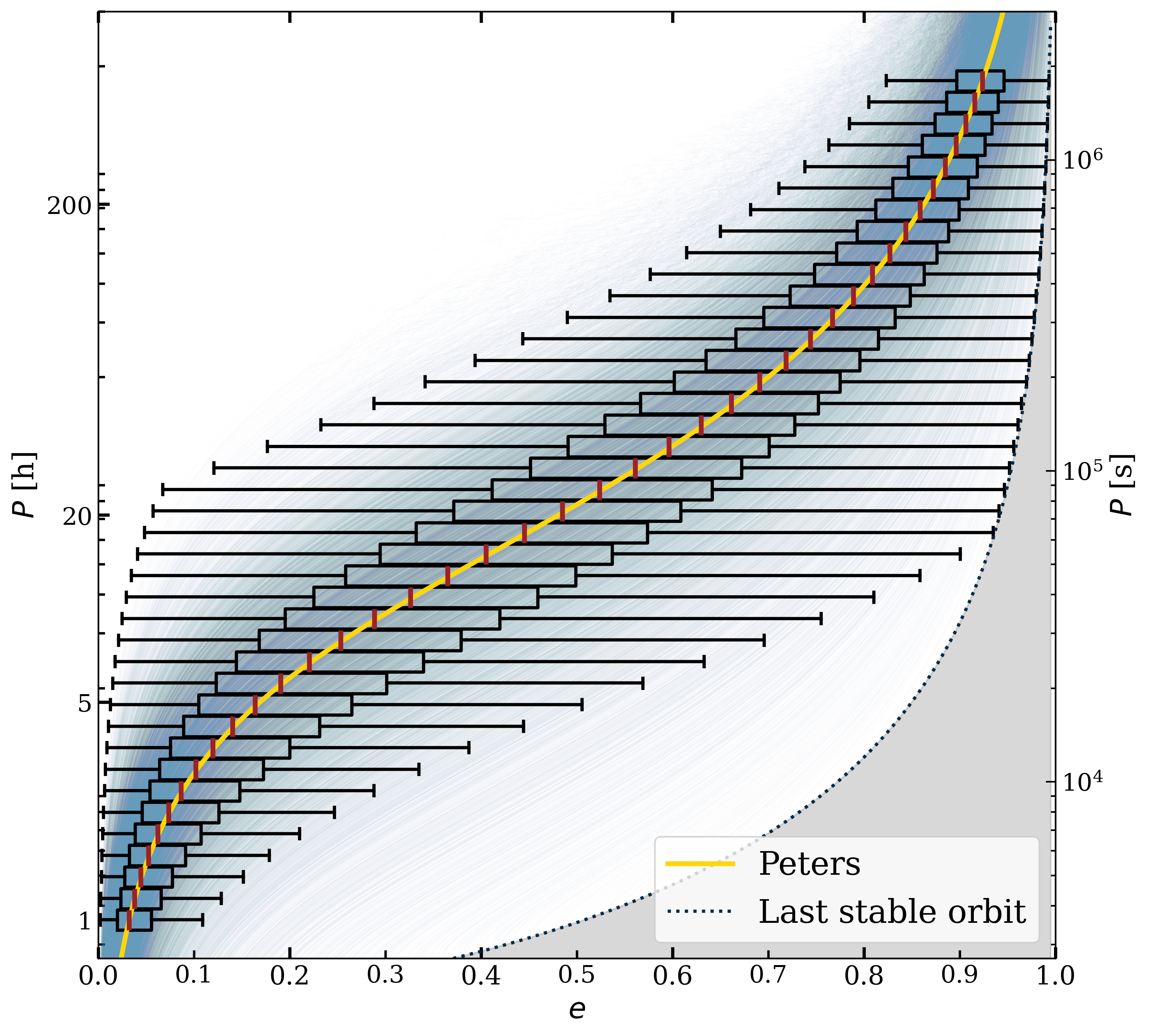}
    \caption{ 
    For each period in a range from $1-500$h, we show the corresponding PDF using a box and whisker representation.
    The red central line in each box indicates the median of the corresponding distribution, while the box contains the central $50$ per cent (interquartile range, IQR) and the whiskers extend from the lower and upper quartiles to the most extreme values within 1.5 IQR of each quartile. Data points lying beyond these limits are treated as outliers and not plotted. The yellow solid line represent the analytical evolution through Peters equation. The background solid light-blue lines are the trajectories of the various systems starting at different initial periods until plunge. The dotted blue line marks the last stable orbit, i.e. the separatrix. The gray shaded area below it is the loss cone. }
    \label{fig:QPE}
\end{figure}

Figure \ref{fig:QPE} provides such a diagnostic for the Hills population considered here. At a given orbital period, it shows the range of eccentricities reached by our systems during their GW-driven evolution. The period range has been chosen to cover the typical $1-200$h range for the QPE recurrence time, assuming a model with one or two piercings per orbit.
For example, \citet{Franchini2023} inferred a mean eccentricity of approximately $\bar{e}\simeq 0.4$ for the QPE source RXJ1301, modelling it as an EMRI with a period of about six hours. This lies within the region occupied by our Hills population and would therefore make a Hills origin a hypothesis worth exploring for this source. However, this should not be interpreted as evidence for a Hills origin, since the inferred orbital parameters are model-dependent and in this case the source itself presents significant modelling challenges, making its eccentricity measurements unreliable.

We point out that the median points closely follow Peters equations \citep{Peters1964, Maggiore2007} (with initial conditions set to ensure the solution passes through the median of the PDF for the largest period)
as they are still far from the separatrix at these orbital period, and will instead deviate from the yellow curve when closer to the separatrix, at smaller periods.

More generally, current disc-crossing models often favour nearly circular or weakly eccentric orbits for QPEs. If very low eccentricities are required, the Hills channel might reproduce only a subset of the observed parameter space  \citep[see also,][]{ZhouZhong2024}.
Moreover, \citet{Allievi2026} show that realistic EMRI eccentricity distributions and event rates pose significant challenges for explaining the overall QPE population with two-body relaxation EMRIs alone. Our results therefore do not establish the Hills mechanism as a general explanation for QPEs. Instead, Fig. \ref{fig:QPE} provides a physically motivated diagnostic that can be applied as more reliable, model-independent constraints on QPE orbital parameters become available in the future.

\subsection{Conclusion}
In summary, our calculations show that the Hills mechanism can produce a distinct population of EMRIs that is characterized by lower eccentricities than classical relaxation-driven EMRIs and by a potentially significant contribution to the EMRI population if BBHs are sufficiently abundant in NSCs. The combination of eccentricity measurements and population information therefore offers a promising way to constrain the origin of EMRIs with LISA. At the same time, the strong dependence of the rates and orbital distributions on the assumed BBH and NSC properties highlights the need for a broader population study before these results can be generalized to the cosmic EMRI population.

%%%%%%%%%%%%%%%%%%%%%%%%%%%%%%%%%%%%%%%%%%%%%%%%%%%%%%%%%%%%%%
\section*{Data Availability}
The data underlying this article will be shared on reasonable request to the corresponding author.
Part of the data used in this work is publicly available at \url{https://zenodo.org/records/17882705}.

\section*{acknowledgements}
BS and EMR acknowledge support from European Research Council (ERC) grant number:
101002511/project acronym: VEGA P.
Part of this work was supported by \emph{ESO}, project number Ts~17/2--1.
DM acknowledges that this publication was produced while attending the PhD program in Space Science and Technology at the University of Trento, Cycle XXXIX, with the support of a scholarship financed by the Ministerial Decree no.\ 118 of 2nd March 2023, based on the NRRP - funded by the European Union - NextGenerationEU - Mission 4 ``Education and Research'', Component 1 ``Enhancement of the offer of educational services: from nurseries to universities'' - Investment 4.1 ``Extension of the number of research doctorates and innovative doctorates for public administration and cultural heritage'' - CUP E66E23000110001.
AS acknowledges the financial support provided under the European Union’s H2020 ERC Advanced Grant ``PINGU'' (Grant Agreement: 101142079).

%------------------------------------------------------------------
\vspace{-1.0em}
\bibliographystyle{mnras}
\bibliography{bibliography}

%%%%%%%%%%%%%%%%%%%%%%%%%%%%%%%%%%%%%%%%%%%%%%%%%%%%
\appendix
\section{Error bars}\label{appendix}
    Error bars $\Delta \mathcal{P}$ have been computed by propagation of uncertainty as
\begin{equation}
\begin{aligned}
\Delta \mathcal{P}^2&=\left(\frac{\partial \mathcal{P}}{\partial H}\right)^2(\Delta H)^2+\left(\frac{\partial \mathcal{P}}{\partial C}\right)^2(\Delta C)^2\\
&=\frac{C^2(\Delta H)^2+H^2(\Delta C)^2}{(H+C)^4}
\end{aligned}
\end{equation}
where $H$ and $C$ are the PDFs for the Hills and classical channel respectively
$\Delta C$ and $\Delta H$ are computed as $\Delta H_i=\sqrt{\sum_{j\in i}(w_j)^2}$, based on the weights $w_j$ in each $i$-th bin of the corresponding histogram.
We used that
\begin{equation}
 \frac{\partial \mathcal{P}}{\partial H}=\frac{C}{(H+C)^2}, \quad \frac{\partial \mathcal{P}}{\partial C}=-\frac{H}{(H+C)^2}.
\end{equation}
Eventually, error bars are computed as
\begin{equation}
\Delta \mathcal{P}=\frac{\sqrt{C^2(\Delta H)^2+H^2(\Delta C)^2}}{(H+C)^2}.
\end{equation}

\bsp
\label{lastpage}
\end{document}